\documentclass[conference,twoside]{IEEEtran}

\usepackage{iftex}
\ifPDFTeX
  \PackageError{noriega2026}{Compile this manuscript with XeLaTeX}{The embedded PGFPlots typography requires fontspec/unicode-math so plot text uses Fira Sans and plot math uses Fira Math. pdfLaTeX cannot load OpenType Fira Math and will produce serif plot math.}
  \usepackage[T1]{fontenc}
  \usepackage{newpxtext}
  \usepackage{newpxmath}
  \usepackage{FiraSans}
  \usepackage[scaled=0.92]{beramono}
  \newcommand{\bodyfont}{\rmfamily}
  \newcommand{\figfont}{\sffamily}
  \DeclareRobustCommand{\plotmathversion}{}
\else
  \usepackage{fontspec}
  \newfontfamily\bodyfont{texgyrepagella-regular.otf}[
    Ligatures=TeX,
    BoldFont=texgyrepagella-bold.otf,
    ItalicFont=texgyrepagella-italic.otf,
    BoldItalicFont=texgyrepagella-bolditalic.otf]
  \newfontfamily\figfont{FiraSans-Regular.otf}[
    Ligatures=TeX,
    BoldFont=FiraSans-Bold.otf,
    ItalicFont=FiraSans-Italic.otf,
    BoldItalicFont=FiraSans-BoldItalic.otf]
  \usepackage{unicode-math}
  \DeclareRobustCommand{\plotmathversion}{\mathversion{plotsans}}
\fi

\usepackage{graphicx}
\ifPDFTeX
  \usepackage{amsmath,amssymb}
\else
  \usepackage{amsmath}
\fi
\usepackage{booktabs}
\usepackage{tabularx}
\usepackage{array}
\usepackage{xcolor}
\usepackage{hyperref}
\hypersetup{
  colorlinks=true,
  linkcolor=blue!50!black,
  citecolor=blue!50!black,
  urlcolor=blue!50!black,
}
\DeclareRobustCommand{\monourl}[1]{{\ttfamily\href{#1}{\nolinkurl{#1}}}}
\usepackage{caption}
\usepackage{adjustbox}
\usepackage{placeins}
\usepackage{stfloats}
\usepackage{etoolbox}
\usepackage{tikz}
\usepackage{pgfplots}
\usepgfplotslibrary{groupplots,fillbetween}
\usetikzlibrary{calc}
\pgfplotsset{compat=1.18}
\usepackage{fancyhdr}

\definecolor{orcidgreen}{HTML}{A6CE39}
\DeclareRobustCommand{\orcidlink}[1]{%
  \href{https://orcid.org/#1}{%
    \tikz[baseline=-0.45ex]{%
      \fill[orcidgreen] (0,0) circle (0.82ex);
      \node[white,font=\sffamily\bfseries\tiny,inner sep=0pt] at (0,0) {iD};
    }%
  }%
}

\renewcommand{\footnoterule}{%
  \kern-3pt
  \hrule width 0.35\columnwidth height 0.4pt
  \kern 2.6pt
}

\makeatletter
\long\def\@makefntext#1{%
  \parindent 1em\noindent
  {\normalfont\figfont\plotmathversion\scriptsize
   \hb@xt@1.8em{\hss\@makefnmark}#1}}
\makeatother

\makeatletter
\newif\ifFloatStackHasFigure
\newif\ifFloatStackHasTable
\newcommand{\ScanFloatStackTypes}[1]{%
  \global\FloatStackHasFigurefalse
  \global\FloatStackHasTablefalse
  \begingroup
    \let\@elt\CheckFloatStackType
    #1%
  \endgroup
}
\newcommand{\CheckFloatStackType}[1]{%
  \@tempcnta\count#1\relax
  \divide\@tempcnta\@xxxii
  \ifnum\@tempcnta=\ftype@figure\relax
    \global\FloatStackHasFiguretrue
  \fi
  \ifnum\@tempcnta=\ftype@table\relax
    \global\FloatStackHasTabletrue
  \fi
}

\newcommand{\DoubleFloatTopBoundaryRule}{%
  \ifFloatStackHasFigure
    \vskip 2pt plus 0.5pt
    \hrule width \textwidth height 0.4pt
  \fi
}
\newcommand{\DoubleFloatBottomBoundaryRule}{%
  \ifFloatStackHasFigure
    \hrule width \textwidth height 0.4pt
    \vskip 2pt plus 0.5pt
  \else\ifFloatStackHasTable
    \hrule width \textwidth height 0.4pt
    \vskip 2pt plus 0.5pt
  \fi\fi
}

\def\dblfigrule{\hrule width \textwidth height 0.4pt}
\pretocmd{\@cflt}{\ScanFloatStackTypes\@toplist}{}{}
\pretocmd{\@cflb}{\ScanFloatStackTypes\@botlist}{}{}
\pretocmd{\@cdblflt}{\ScanFloatStackTypes\@dbltoplist}{}{}
\pretocmd{\@cdblflb}{\ScanFloatStackTypes\@dblbotlist}{}{}
\patchcmd{\@cdblflt}{\dblfigrule}{\DoubleFloatTopBoundaryRule}{}{}
\patchcmd{\@cdblflb}{\dblfigrule}{\DoubleFloatBottomBoundaryRule}{}{}
\makeatother
\newcommand{\floatcaprule}{}
\newcommand{\figcaprule}{}
\newcommand{\tabnoterule}{%
  \par\vspace{1pt}%
  \hrule width \linewidth height 0.4pt%
}
\renewcommand{\abstractname}{\bodyfont Abstract}
\makeatletter
\def\abstract{%
  \normalfont\bodyfont\bfseries\small
  \par\indent{\figfont\bfseries\itshape\abstractname}---\normalfont\bodyfont\bfseries\small\ignorespaces}

\makeatother

\fancypagestyle{plain}{%
  \fancyhf{}
  \fancyfoot[C]{\scriptsize\thepage}
  
}
\pgfplotsset{
  colormap={viridisWarmSaturation}{
    rgb255(0cm)=(255,247,228);
    rgb255(25cm)=(248,213,167);
    rgb255(50cm)=(235,166,104);
    rgb255(75cm)=(216,112,55);
    rgb255(100cm)=(188,62,29)
  },
  colormap={orangeRedSequential}{
    rgb255(0cm)=(255,248,235);
    rgb255(25cm)=(252,210,151);
    rgb255(50cm)=(244,156,84);
    rgb255(75cm)=(216,89,54);
    rgb255(100cm)=(155,38,60)
  },
  colormap={viridisBlueGreen}{
    rgb255(0cm)=(247,252,250);
    rgb255(25cm)=(203,235,230);
    rgb255(50cm)=(132,205,198);
    rgb255(75cm)=(62,160,151);
    rgb255(100cm)=(33,145,140)
  },
}


\DeclareCaptionFont{figsans}{\figfont\plotmathversion}
\AtBeginDocument{%
  \setlength{\abovedisplayskip}{2pt plus 1pt minus 1pt}%
  \setlength{\belowdisplayskip}{2pt plus 1pt minus 1pt}%
  \setlength{\abovedisplayshortskip}{2pt plus 1pt minus 1pt}%
  \setlength{\belowdisplayshortskip}{2pt plus 1pt minus 1pt}%
  \setlength{\skip\footins}{8pt plus 2pt minus 1pt}%
  \setlength{\footnotesep}{3pt}%
}
\newcolumntype{Y}{>{\raggedright\arraybackslash}X}

\definecolor{condBaseline}{HTML}{444444}
\definecolor{condAugOne}{HTML}{7B1E3A}
\definecolor{condAugTwo}{HTML}{2D708E}
\definecolor{condFiltOne}{HTML}{20A387}
\definecolor{condFiltTwo}{HTML}{73D055}
\definecolor{classGlioma}{HTML}{440154}
\definecolor{classMeningioma}{HTML}{2D708E}
\definecolor{classNoTumour}{HTML}{20A387}
\definecolor{classPituitary}{HTML}{73D055}

\tikzset{
  boxAugOne/.style={draw=black, fill=condAugOne, line width=0.58pt},
  boxAugTwo/.style={draw=black, fill=condAugTwo, line width=0.58pt},
  boxFiltOne/.style={draw=black, fill=condFiltOne, line width=0.58pt},
  boxFiltTwo/.style={draw=black, fill=condFiltTwo, line width=0.58pt},
  boxSigAugOne/.style={draw=black, fill=condAugOne, line width=1.05pt},
  boxSigAugTwo/.style={draw=black, fill=condAugTwo, line width=1.05pt},
  boxSigFiltOne/.style={draw=black, fill=condFiltOne, line width=1.05pt},
  boxSigFiltTwo/.style={draw=black, fill=condFiltTwo, line width=1.05pt},
  whiskerAugOne/.style={draw=black, line width=0.48pt},
  whiskerAugTwo/.style={draw=black, line width=0.48pt},
  whiskerFiltOne/.style={draw=black, line width=0.48pt},
  whiskerFiltTwo/.style={draw=black, line width=0.48pt},
  whiskerSig/.style={draw=black, line width=0.58pt},
  meanAugOne/.style={draw=black, line width=0.72pt},
  meanAugTwo/.style={draw=black, line width=0.72pt},
  meanFiltOne/.style={draw=black, line width=0.72pt},
  meanFiltTwo/.style={draw=black, line width=0.72pt},
  meanSig/.style={draw=black, line width=0.72pt}
}

\newcommand{\plotfigurefont}{\figfont\plotmathversion}

\newcommand{\metriccell}[2]{#1~\textpm{}~#2}

\makeatletter
\def\section{\@startsection{section}{1}{\z@}{1.5ex plus 1.5ex minus 0.5ex}%
{0.7ex plus 1ex minus 0ex}{\normalfont\normalsize\figfont\scshape\bfseries\centering}}
\def\subsection{\@startsection{subsection}{2}{\z@}{1.5ex plus 1.5ex minus 0.5ex}%
{0.7ex plus .5ex minus 0ex}{\normalfont\normalsize\bodyfont\bfseries\raggedright}}
\def\subsubsection{\@startsection{subsubsection}{3}{\z@}{1.5ex plus 1.5ex minus 0.5ex}%
{0.5ex plus .5ex minus 0ex}{\normalfont\normalsize\bodyfont\raggedright}}
\def\paragraph{\@startsection{paragraph}{4}{2\parindent}{0ex plus 0.1ex minus 0.1ex}%
{0ex}{\normalfont\normalsize\bodyfont\itshape}}
\def\subparagraph{\@startsection{subparagraph}{5}{2\parindent}{0ex plus 0.1ex minus 0.1ex}%
{0ex}{\normalfont\normalsize\bodyfont\itshape}}
\makeatother

\AtBeginDocument{\bodyfont}

\begin{document}

\title{\sffamily\bfseries
\begin{minipage}{0.92\textwidth}
\centering
\rule{\linewidth}{0.35pt}\par\vspace{0.9ex}
Do Synthetic Brain MRIs Reliably Improve\\ Tumour Classification?\\ A StyleGAN2-ADA Class-Plane\\ Augmentation Study on BRISC 2025
\par\vspace{-0.20\baselineskip}
\rule{\linewidth}{0.35pt}
\end{minipage}}

\author{\bodyfont\IEEEauthorblockN{\bodyfont\large Jos\'e Rafael Noriega Cede\~no\,\orcidlink{0009-0003-5258-9611}}
\IEEEauthorblockA{\bodyfont Faculty of Applied Sciences and Technology\\
Humber Polytechnic\\
Toronto, ON M9W 5L7\\
\href{mailto:n10007954@humber.ca}{n10007954@humber.ca}}}

\maketitle
\thispagestyle{plain}
\bodyfont

\begin{abstract}
\bodyfont\bfseries\small
Generative augmentation is often proposed as a remedy for small medical-image datasets, but synthetic images are only useful when they improve downstream task performance. ``Augmentation'' here means synthetic supplementation: GAN-generated samples added to the real training pool, not geometric or photometric transforms of existing images. Twelve class-plane StyleGAN2-ADA generators were trained on constrained BRISC 2025 partitions to test whether their output, with or without InceptionV3 feature-space filtering, improves held-out tumour classification across three classifier families: a random forest (RF) on InceptionV3 features, a compact two-headed convolutional neural network (CNN), and MobileViTV2, a mobile hybrid convolutional--transformer. Each was evaluated at 1:1 and 1:2 real-to-synthetic ratios. An independent GPT-5.5 blind test placed gated real-versus-synthetic discrimination at 57.73\% (95\% CI: 54.48--60.92\%) on the model-legible subset---modestly above chance. The RF classifier did not benefit from the synthetic MRIs. The CNN showed consistent mean gains that did not survive Holm correction. MobileViTV2 showed the clearest benefit: filtered 1:1 augmentation improved tumour classification accuracy by 1.02\% absolute (95\% CI: 0.54--1.54\%; Holm-corrected \textit{p} = 0.0104). A secondary efficiency analysis found that every augmented CNN condition selected its checkpoint 42--64\% earlier than baseline, while compute-matched MobileViTV2 runs reached selection after 50--67\% fewer real-data epochs. Overall, augmentation utility was found to be architecture- and ratio-dependent, not guaranteed by visual fidelity alone.
\end{abstract}
\noindent{\normalfont\bodyfont\bfseries\small\indent{\bodyfont\small\itshape\bfseries Keywords}\,---\,brain tumour MRI, BRISC 2025, StyleGAN2-ADA, synthetic augmentation, RF, CNN,  MobileViTV2, medical image synthesis.}
\par\addvspace{0.5\baselineskip}
\bodyfont\small

\section{Introduction}

Generative adversarial networks (GANs) shifted the practical question of image synthesis from whether artificial images can be produced to whether they can be made coherent enough to support scientific or operational use. Since the original GAN formulation \cite{goodfellow2014gan}, convolutional adversarial models such as DCGAN \cite{radford2016dcgan} and later style-based generators have progressively improved visual fidelity, controllability, and training stability. StyleGAN2 reduced several characteristic artifacts of earlier style-based synthesis \cite{karras2020stylegan2}, and StyleGAN2-ADA introduced adaptive discriminator augmentation to reduce overfitting when the real training set is small \cite{karras2020ada}. StyleGAN3 later introduced alias-free design choices \cite{karras2021stylegan3}, but this project uses the StyleGAN2 architecture and loss with ADA, launched through the NVIDIA StyleGAN3 PyTorch codebase.

\begin{figure}[!tbp]
  \floatcaprule
  \centering
  \includegraphics[width=\linewidth,height=0.55\textheight,keepaspectratio]{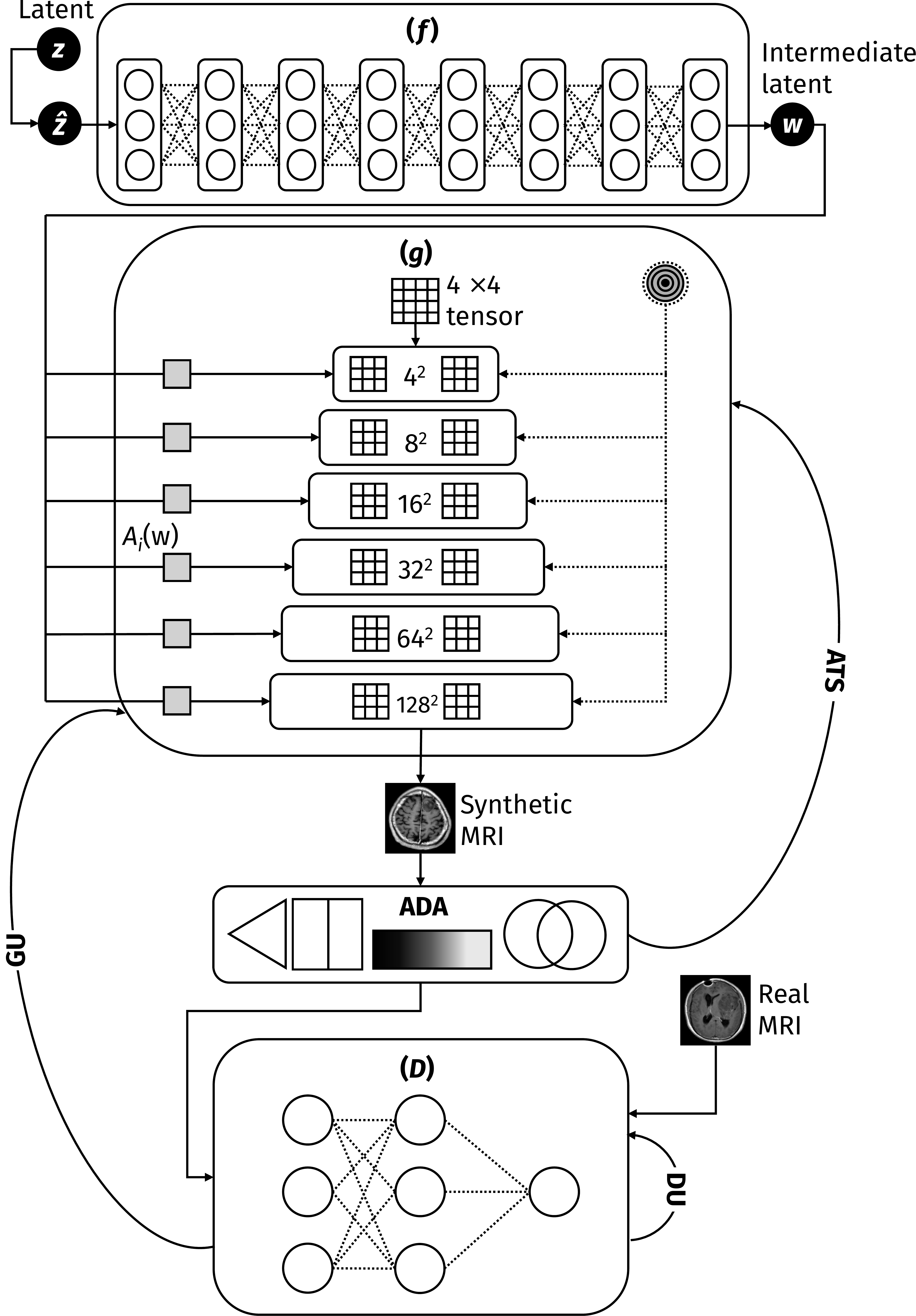}
  \caption{StyleGAN2-ADA training loop used for class-plane brain MRI synthesis. The label-free schematic is read from top to bottom: a latent vector $z$ is normalized to $\hat z$, passed through the eight-layer mapping network $f$, and converted to the intermediate latent $w$. The synthesis network $g$ starts from a learned $4\times4$ constant, then passes through six progressively larger synthesis blocks; each block receives a style transform of $w$, a per-layer noise input, and two modulated/demodulated convolution operations before producing a synthetic MRI. The transform bar below the generated slice represents adaptive discriminator augmentation (ADA), while the discriminator compares generated and real MRI inputs and sends adversarial feedback to the generator. The complete process was repeated for twelve independent generators, one for each tumour-class by anatomical-plane combination.}
  \label{fig:stylegan-diagram}
  \figcaprule
\end{figure}

In medical imaging the limited-data problem is more than a technical curiosity, but an actual structural constraint: datasets are bounded by privacy, institutional access, annotation cost, disease rarity, acquisition heterogeneity, and class imbalance. In brain MRI the constraint sharpens once images are divided not only by diagnosis but also by anatomical plane; a dataset that looks moderately sized in aggregate can become small as soon as clinically meaningful tumour-plane subsets are isolated.

Synthetic augmentation is attractive in this setting, but the word needs to be used \textit{carefully}. Here, it refers to synthetic supplementation, or generative oversampling: adding GAN-synthesized images alongside real training images. It does not refer to classical data augmentation, where an existing image is jittered, flipped, cropped, or otherwise transformed. The scientific question, then, is not whether GANs can produce realistic-looking images, but whether the synthesized images add useful variation for a downstream classifier evaluated on real held-out data. After all, visual realism, Fr\'echet Inception Distance (FID) \cite{heusel2017ttur}, and other generic image-quality metrics do not always track diagnostic or algorithmic utility in medical tasks. Synthetic images can help by increasing apparent diversity, but they can also hurt by introducing artifacts, duplicate-like patterns, or distribution shifts that classifiers learn as shortcuts.

Accordingly, the study asks three linked questions. First, can StyleGAN2-ADA learn plausible tumour-plane BRISC MRI distributions from only a few hundred real images per generator? Second, do the generated images improve held-out tumour classification when added to the training set? Third, do synthetic images change the training process itself, reducing seed-to-seed volatility and the training effort required to reach the selected checkpoint, even when final accuracy moves only modestly? To answer them, twelve separate StyleGAN2-ADA models were trained (Fig.~\ref{fig:stylegan-diagram}), synthetic MRI images were generated and optionally filtered in InceptionV3 feature space, and downstream performance, seed dispersion, and training-history dynamics were compared across RF, CNN, and MobileViTV2 classifiers under multiple augmentation ratios.

\section{Related Work}

GAN-based medical image synthesis has a substantial history. Yi, Walia, and Babyn surveyed adversarial learning across synthesis, reconstruction, segmentation, and domain adaptation in medical imaging \cite{yi2019ganreview}; Frid-Adar \textit{et al.} provided an early demonstration that GAN-generated images could improve CNN classification of CT liver lesions \cite{fridadar2018livergan}; and Shin \textit{et al.} and Han \textit{et al.} established that synthetic brain MRI generation is feasible \cite{shin2018synthesis,han2018brainmri}. These results show the technique is viable, but feasibility and utility are different questions.

More recent work narrows the gap. Mukherkjee \textit{et al.} developed AGGrGAN, an aggregation of two DCGAN variants and a WGAN with style transfer, and evaluated brain-tumour MRI synthesis using image-similarity measures including SSIM and PSNR \cite{mukherjee2022aggrgan}. Woodland \textit{et al.} applied StyleGAN2-ADA primarily to other medical modalities and included a public brain-tumour dataset as a supplementary experiment, achieving strong FID values under data conditions substantially larger than the per-generator BRISC partitions used here \cite{woodland2022stylegan2ada}. The most directly relevant precedent is Akbar \textit{et al.}, who compared progressive GAN, all three StyleGAN versions, and diffusion models for brain-tumour MRI segmentation and reported that FID and Inception Score did not reliably predict downstream task performance \cite{akbar2024syntheticmri}. That finding is the \textit{immediate} motivation for measuring augmentation value through held-out classification rather than image-quality metrics alone.

Diffusion models have emerged as a serious synthesis alternative. Pinaya \textit{et al.} generated high-resolution brain images with latent diffusion \cite{pinaya2022brainldm}, Khader \textit{et al.} extended denoising diffusion to 3D medical volumes \cite{khader2023ddpmmedical}, and M{\"u}ller-Franzes \textit{et al.} compared latent diffusion and GAN approaches across modalities \cite{mullerfranzes2023multimodal}. StyleGAN2-ADA was chosen here because it offers a well-established limited-data training recipe, fits twelve independent two-dimensional generators within a single-GPU budget, and has direct medical-imaging precedent. A diffusion comparison under the same class-plane constraints remains an open question.

The dataset itself has relevant supervised baselines. BRISC is a public dataset for brain-tumour segmentation and classification, with 6,000 contrast-enhanced T1-weighted MRI scans for classification split into 5,000 training and 1,000 test images \cite{fateh2026brisc}. Thahiruddin and Wulandari compared ResNet18, EfficientNet-B0, MobileNetV3-Small, and MobileViTV2 on BRISC \cite{thahiruddin2025brisc}. Ali and Behzad separately combined DC-GAN brain-MRI synthesis with a compact CNN classifier using 128~$\times$~128 grayscale images, convolutional blocks, global average pooling, and a 1024-unit dense layer \cite{ali2025dcgancnn}. These works establish that brain-tumour MRI can support supervised CNN and hybrid-transformer classifiers, but none of them evaluates StyleGAN2-ADA class-plane generators, feature-space filtering, paired multi-seed augmentation utility, or training-efficiency effects under constrained BRISC partitions.

What is new here is the combination: BRISC 2025 as a multi-class tumour classification benchmark; twelve small class-plane StyleGAN2-ADA generators; both unfiltered and diversity-filtered augmentation at two real-to-synthetic ratios; evaluation spanning a fixed-feature classical baseline, a convolutional model, and a hybrid transformer; and paired multi-seed testing with family-wise correction. The aim is not to show that synthetic MRI can be made, but to establish the conditions under which it is actually useful.

\section{Materials and Methods}

\subsection{Computing Resources}

All experiments were conducted on a Windows workstation equipped with a single NVIDIA GeForce RTX 5070 GPU (12.8 GB-class VRAM). The software environment comprised \texttt{Python 3.13.13}, \texttt{PyTorch 2.12.0.dev20260221+cu128}\footnote{A pre-release CUDA~12.8 nightly wheel, not a stable PyTorch release.}, \texttt{CUDA 12.8}, \texttt{scikit-learn 1.8.0}, \texttt{NumPy 2.4.2}, \texttt{SciPy 1.17.0}, \texttt{Pillow 12.1.0}, and \texttt{timm 1.0.25}. The PyTorch wheel is reported explicitly because it was a nightly build required for the local CUDA stack; exact numerical replication may require the same dated wheel or the closest archived nightly. Each StyleGAN2-ADA generator was trained with total batch size $B=16$ and per-GPU microbatch $B_{\mathrm{GPU}}=4$,\footnote{With \texttt{--batch=16} and \texttt{--batch-gpu=4}, discriminator gradients are accumulated over four sub-batches before each optimizer update, allowing a larger effective batch size on the local single-GPU setup.} R1 regularization $\gamma = 2$, and a budget of $K=1{,}000$ kimg per tumour-plane run.

The hardware shaped several design choices. Training twelve independent generators is expensive, but each one then learns a narrower, more coherent distribution; the trade-off against reduced per-generator sample size is one of the questions this experiment is built to evaluate. The single-GPU memory ceiling fixed the training resolution at 128~$\times$~128 pixels and bounded the range of augmentation ratios tested downstream. Completed 1,000-kimg StyleGAN2-ADA logs ran from roughly 9.8 to 27.6 wall-clock hours per tumour-plane run, totalling approximately 162 generator GPU-hours across the twelve final runs. Downstream classifier and VLM-audit costs were not used as optimization criteria. MobileViTV2 augmented and baseline runs were matched by optimizer step count so that augmented conditions did not benefit simply from having more data to iterate over.

\subsection{Dataset Acquisition and Integrity}

The dataset was obtained from the official BRISC 2025 Kaggle release \cite{fateh2025kaggle} by Fateh \textit{et al.} \cite{fateh2026brisc}. The public classification package contains 6,000 contrast-enhanced T1-weighted MRI images, with 5,000 assigned to training and 1,000 assigned to test. The four diagnostic classes are \textit{glioma}, \textit{meningioma}, \textit{pituitary tumour}, and \textit{no tumour}; each image is also labelled by \textit{axial}, \textit{coronal}, or \textit{sagittal} plane.

This study used only publicly released, de-identified BRISC images and did not access patient identifiers or protected health information. No new human-subject data were collected, so institutional review was not required for this secondary analysis.

The classification split was read directly from the BRISC directory structure. Diagnostic class was taken from the class folder, and anatomical plane was parsed from the filename suffixes (\texttt{ax}, \texttt{co}, \texttt{sa}). This distinction mattered because every generative model was trained at the class-plane level rather than at the class level alone. The training data were consequently reorganized into twelve tumour-plane subsets before StyleGAN2-ADA training, while the downstream classifier manifests retained both the tumour and plane labels for the two-headed models.

A project-side audit detected exact train-test image overlap: 104 training images that also appeared in the held-out test set were removed from the classifier training manifest. Removal was conservative and image-based; each candidate was compared against test images by pixel hash and then confirmed by exact array equality before being dropped. The removals were not class-balanced: 97 were meningioma and 7 pituitary; by plane, 64 were axial, 24 coronal, and 16 sagittal, with most duplicates concentrated in meningioma axial images (62/104). The final classifier training set contained 4,896 real images and the held-out test set 1,000 (Table~\ref{tab:dataset-composition}). After removal, the audit found no filepath, basename, file SHA256, pixel-exact, or pixel-hash-exact overlaps between the final train and test manifests. The same audit flagged 356 close perceptual-hash neighbours\footnote{Perceptual-hash proximity was flagged at a Hamming distance threshold of $\leq 6$ bits on 64-bit DCT-based perceptual hashes (pHash \cite{zauner2010phash}, computed via \texttt{imagehash.phash} with default 8~$\times$~8 hash size).} and 5 close Inception-feature neighbours,\footnote{Inception-feature proximity was flagged when the cosine distance between InceptionV3 pool3 embeddings of a train and a test image fell below 0.01 (i.e., cosine similarity $> 0.99$), indicating near-identical feature representations.} which were retained as diagnostic warnings rather than treated as exact duplicates. No patient-identifier columns were available, so the audit guarantees image-level non-overlap (and as such, prevented a plausible instance of data leakage) but cannot establish patient-level independence.

\begin{table}[!tbp]
  \floatcaprule
\caption{Final real-data composition used by the downstream classifiers after exact train-test overlap removal. The upper block reports real training counts by tumour class and anatomical plane after the 104 duplicate training images were removed; the rightmost column gives the class totals used to set synthetic augmentation quotas. The lower block reports the unchanged 1,000-image held-out BRISC test set under the same class-plane structure, providing the denominators for both the tumour and plane heads.}
\label{tab:dataset-composition}
\centering
\normalfont\plotfigurefont\scriptsize
\begin{tabularx}{\linewidth}{Yrrrr}
\toprule
Class & Axial & Coronal & Sagittal & Train total \\
\midrule
Glioma & 394 & 430 & 323 & 1,147 \\
Meningioma & 361 & 407 & 464 & 1,232 \\
No tumour & 352 & 310 & 405 & 1,067 \\
Pituitary & 424 & 505 & 521 & 1,450 \\
\midrule
Total & 1,531 & 1,652 & 1,713 & 4,896 \\
\bottomrule
\end{tabularx}
\vspace{0.2em}
\begin{tabularx}{\linewidth}{Yrrrr}
\toprule
Held-out test class & Axial & Coronal & Sagittal & Test total \\
\midrule
Glioma & 85 & 81 & 88 & 254 \\
Meningioma & 137 & 86 & 83 & 306 \\
No tumour & 52 & 48 & 40 & 140 \\
Pituitary & 124 & 90 & 86 & 300 \\
\midrule
Total & 398 & 305 & 297 & 1,000 \\
\bottomrule
\end{tabularx}
\end{table}

\subsection{Preprocessing and Harmonization}

\begin{figure}[!tbp]
  \floatcaprule
  \centering
  \includegraphics[width=0.79\linewidth]{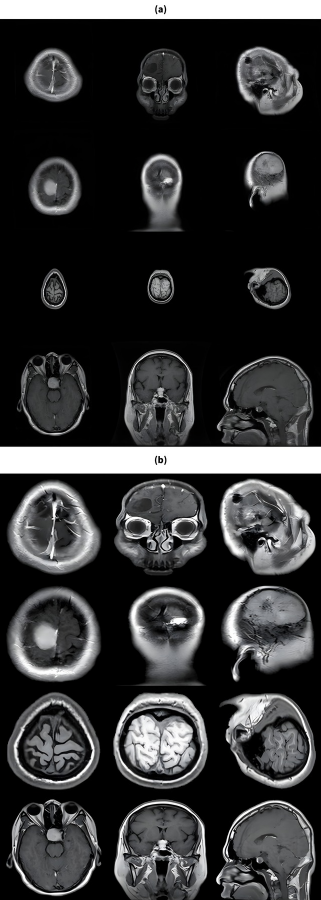}
  \caption{Raw-to-preprocessed BRISC examples documenting the harmonization applied before GAN training and downstream classification. Panel \textbf{(a)} shows raw MRI examples arranged as tumour class by anatomical plane, exposing the original variation in field of view, skull visibility, background intensity, crop position, and contrast. Panel \textbf{(b)} shows the same images after grayscale conversion, brain-region masking, percentile-based intensity normalization, foreground cropping, resizing to 128~$\times$~128 pixels, adaptive sharpening, remasking, and lossless PNG export. The paired layout makes the point that preprocessing was not cosmetic cleanup: it places real and synthetic images in the same standardized visual space before classifiers can learn source-specific shortcuts. Images are displayed here at a resolution higher than 128~$\times$~128 for illustrative purposes; 128~$\times$~128 was the working resolution used for all GAN training, synthetic generation, and downstream classification.}
  \label{fig:raw-preprocessed}
  \figcaprule
\end{figure}

Real and synthetic images were processed through the same deterministic pipeline before classifier training, so that source-specific presentation differences could not become a classifier shortcut (Fig.~\ref{fig:raw-preprocessed}).

Each image was first converted to 8-bit grayscale. Foreground extraction used Otsu thresholding combined with border flood-fill skull stripping, applied across five progressively relaxed threshold scales (Otsu $\times 0.50$, $\times 0.75$, $\times 1.00$, $\times 1.25$, $\times 1.50$, with between five and seven binary-closing iterations each) because raw BRISC images vary widely in field of view, skull visibility, and background brightness. For each attempt, the external background was identified by flood filling from a padded image border, and the brain candidate was taken as the non-background region. The candidate mask was then cleaned by retaining the largest 4-connected component, removing fragments below 500 pixels, and applying a final three-iteration binary closing. A mask was accepted if its area occupied 15\% to 85\% of the field of view; if none of the five attempts fell within that range, the closest-to-50\% candidate was used, and if all attempts were implausible, the full image served as a fallback. This cascade avoids silently discarding valid but difficult scans, such as images with unusually dark skull boundaries, heavy cropping, or atypical contrast.

A read-only telemetry rerun over the raw BRISC classification package showed that the conservative Otsu $\times 0.50$ tier accepted 4,995/5,000 training images (99.90\%) and 998/1,000 test images (99.80\%). The Otsu $\times 0.75$ tier accepted 4 training images and 1 test image, the closest-to-50\% fallback was invoked once in each split, and the full-image fallback was never used. Mean accepted mask coverage was 55.37\% for training and 56.15\% for test images. In practice, then, the cascade acted as a safeguard for rare difficult scans rather than as a frequent hidden transformation.

Intensity normalization was restricted to masked pixels. The $1^{\mathrm{st}}$ and $99^{\mathrm{th}}$ intensity percentiles within the mask served as clipping bounds; the clipped image was then linearly rescaled to 0--255. Background pixels were zeroed at the original resolution \emph{before} any resizing, because interpolating first would smear background intensity into the brain boundary and create a source-identifiable edge signal.

The masked image was cropped to its foreground bounding box with 5\% padding of the larger dimension (minimum 4 pixels) and resized to 128~$\times$~128 using Lanczos interpolation; the mask was resized with nearest-neighbour interpolation to avoid boundary smearing. A scale-adaptive unsharp mask was applied only when the crop required substantial upsampling: no sharpening below a scale factor of 1.2, with radius, percent, and threshold interpolated linearly between (1.0~px, 80\%, 4) and (2.5~px, 200\%, 1) across the 1.2--3.0 range and capped above 3.0. The mask was then reapplied to the resized image, and all pixels below intensity~5 were zeroed to remove residual Lanczos fringe. Outputs were saved as lossless single-channel PNG files.

% Intentionally no FloatBarrier here: later methods text may share the page with Fig. 2.

\subsection{StyleGAN2-ADA Training}

StyleGAN2-ADA training was performed with NVIDIA's official StyleGAN3 PyTorch repository \cite{karras2021stylegan3}, configured with \texttt{--cfg=stylegan2}, which selects the StyleGAN2 generator, discriminator, and adversarial loss. The accurate description is StyleGAN2-ADA trained through the StyleGAN3 codebase.

Twelve independent generators were trained, one per tumour-plane combination. The stratified design narrows each generator's target distribution: a single generator covering all tumour types and planes would have to model multiple anatomical orientations, class morphologies, and acquisition-specific appearances at once, whereas a class-plane generator can specialize. The cost is sample size: each generator sees only a few hundred real images. One aim of the experiment, then, is to see whether StyleGAN2-ADA remains useful under that coherence-versus-sample-size trade-off.

Each tumour-plane subset was packaged as an unlabeled StyleGAN image-folder ZIP at 128~$\times$~128 resolution. Labels were not supplied to StyleGAN because conditioning was achieved structurally: each generator saw only one diagnostic class and one anatomical plane. The StyleGAN2 generator used latent dimension $d_z=512$, intermediate latent dimension $d_w=512$, and $L_{\mathrm{map}}=8$ mapping layers. Both generator and discriminator used channel base $C_{\mathrm{base}}=32{,}768$ and channel cap $C_{\max}=512$; the discriminator used minibatch-standard-deviation group size $g_{\mathrm{mbstd}}=4$. Training used Adam for both networks with learning rate $\eta = 0.002$, moment coefficients $(\beta_1,\beta_2) = (0,0.99)$, and numerical constant $\epsilon_{\mathrm{Adam}} = 10^{-8}$. The loss used R1 regularization with $\gamma = 2.0$, style-mixing probability $p_{\mathrm{mix}}=0.9$, path-length regularization weight $\lambda_{\mathrm{pl}}=2$, and exponential moving average length $\tau_{\mathrm{EMA}}=5$ kimg.

\begin{figure}[!tbp]
  \floatcaprule
  \centering
  \includegraphics[width=\linewidth]{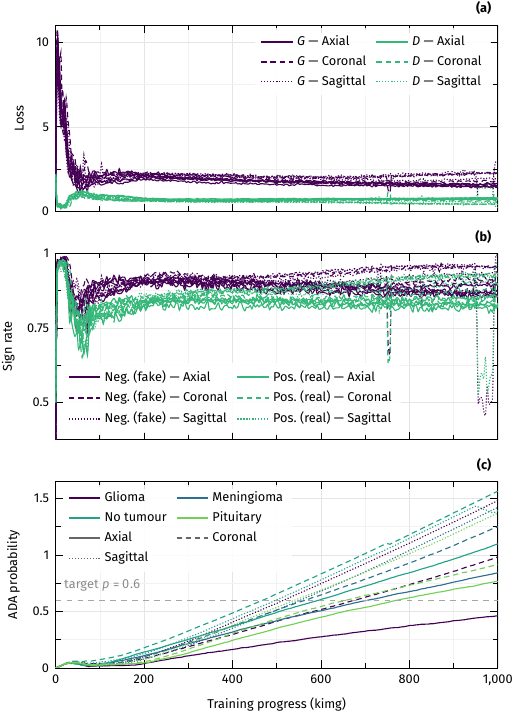}
  \caption{Training dynamics for the twelve class-plane StyleGAN2-ADA generators over 1,000 kimg. In all panels, colour encodes tumour class and line style encodes anatomical plane, so each visible trajectory corresponds to one independent generator. Panel \textbf{(a)} overlays generator and discriminator loss trajectories, summarizing adversarial stability rather than supervised convergence. Panel \textbf{(b)} reports the discriminator real/fake sign statistic; values near 0.5 indicate that the discriminator is uncertain about source assignment, while sustained deviations indicate imbalance between generator and discriminator. Panel \textbf{(c)} reports ADA augmentation probability $p$, which rises when the discriminator begins to overfit the small real subset. High $p$ values accordingly mark generators trained under stronger regularization pressure, a direct consequence of the few-hundred-image class-plane partitions.}
  \label{fig:stylegan-training}
  \figcaprule
\end{figure}

All final generator runs used one GPU, total batch size $B=16$, per-GPU microbatch $B_{\mathrm{GPU}}=4$, total training length $K=1{,}000$ kimg, and $k_{\mathrm{tick}}=4$ kimg per tick. Image and network snapshots were saved every $t_{\mathrm{snap}}=12$ ticks (about 48 kimg intervals). StyleGAN's online metric evaluation was disabled during training to conserve compute; FID, KID, precision, and recall were computed afterwards from saved snapshots. The separation also prevents expensive metric computation from acting as a training-time selection signal. The resulting generator training histories are summarized in Fig.~\ref{fig:stylegan-training}.

\subsection{ADA Augmentation Subset}

The training pipeline intentionally restricted discriminator augmentations. ADA acts inside the discriminator path: real and generated images are transformed before source discrimination, and the generator receives gradients through that transformed fake-image branch. As ADA probability $p$ rises, an augmentation can in turn become part of the effective distribution the generator is rewarded for matching. If the transform preserves the anatomical meaning of the slice, this regularizes the discriminator; if it changes plane semantics, it can leak implausible orientation or preprocessing artifacts into the generator itself.

The saved configurations were thus plane-dependent. In the final \texttt{training\_options.json} files, axial and coronal generators enabled the StyleGAN3 augmentation names \texttt{xflip}, \texttt{brightness}, \texttt{contrast}, \texttt{lumaflip}, and \texttt{imgfilter}. Sagittal generators enabled \texttt{brightness}, \texttt{contrast}, \texttt{lumaflip}, and \texttt{imgfilter}, but not \texttt{xflip}. In axial and coronal views, a horizontal flip is primarily a left--right reflection about the approximate brain midline: it swaps hemisphere laterality but preserves the superior--inferior and anterior--posterior structure of the view, and the tumour label remains valid. For sagittal slices, the horizontal axis carries anterior--posterior information; a horizontal flip can exchange facial/anterior anatomy with occipital/posterior anatomy and teach the discriminator that both profile directions are interchangeable. Once $p$ becomes large, that invariance can encourage the generator to produce mixed or ambiguous sagittal profiles rather than a consistently oriented anatomical plane. Dataset-level mirroring was disabled in all cases; flips, when used, were introduced only through ADA so that their frequency remained adaptive rather than fixed.

The same leakage logic motivated excluding the broader geometric, noise, and cutout components. Rotations, translations, anisotropic scaling, and large resampling changes can make standard MRI planes appear tilted, stretched, off-centre, or partly cropped; at high ADA probability the generator can respond by reproducing those tolerated edge, padding, or plane-mixing artifacts. Injected noise can encourage synthetic grain or speckle unrelated to anatomy, and cutout-like masking can create rectangular defects that are especially problematic in skull-bounded images. Brightness, contrast, luminance inversion, and mild filtering were retained because they mostly vary scanner and preprocessing appearance rather than spatial anatomy, though even these were treated as discriminator regularization, not as claims about physical image acquisition. ADA targeted a discriminator real/fake sign statistic of $r_t=0.6$, raising augmentation strength only when the discriminator began to overfit the small real subset. The restricted, plane-aware ADA subset is a design decision rather than a completed ablation: a matched full-ADA generator family was not trained, so the claim here is anatomical and methodological, not a measured superiority claim for the restricted pipeline.

\subsection{Checkpoint Selection and Image Generation}

StyleGAN2-ADA snapshots were saved periodically during training. Checkpoints were selected through a three-stage audit rather than a pure-FID rule. First, each snapshot was assigned to an FID tier: 0--30 (extraordinary), 30--50 (excellent), 50--75 (good), or above 75 (fair). Only snapshots in the best tier reached by that generator were carried forward. The tier-first ordering was intentionally conservative: it required an acceptable distributional-distance band before precision and recall were used to rank candidates. However, it can exclude a higher-recall snapshot that sits one tier worse on FID, so the low selected-checkpoint recall values reported below should be read partly as a consequence of that rule. Second, snapshots inside the best tier were ranked by a composite score
\begin{equation}
  S = 0.5 \times \text{precision} + 0.5 \times \text{recall}.
\end{equation} The arithmetic mean was used as a ranking aid, not as an F1-like balance claim: because recall was very low across candidate snapshots, a harmonic mean would have collapsed most scores toward zero and offered little practical separation. Third, when the difference between the top two composite scores was $\Delta S \leq 0.005$, recall served as the tie-breaker. This rule was adopted because low FID alone can favour sharp but narrow generators, whereas augmentation utility needs both image quality and adequate coverage of the real distribution.

Each selected snapshot was then evaluated with FID, KID, precision, and recall (Table~\ref{tab:gan-selected-metrics}). KID, or Kernel Inception Distance, estimates the squared maximum mean discrepancy between real and generated Inception feature distributions using a cubic polynomial kernel \cite{binkowski2018mmd}. Unlike FID, which compares Gaussian fits to the feature distributions, KID is a kernel two-sample statistic with an unbiased finite-sample estimator. The StyleGAN3 implementation used here computes \texttt{kid50k\_full} from 50,000 generated images, all available real images in the corresponding class-plane subset, and 100 random subsets of at most 1,000 feature vectors. Thus, in this project, ``50k'' refers to generated samples only; the real side used the full available BRISC subset, which ranged from 310 to 521 real images.

\begin{figure}[!tbp]
  \floatcaprule
  \centering
  \includegraphics[width=\linewidth]{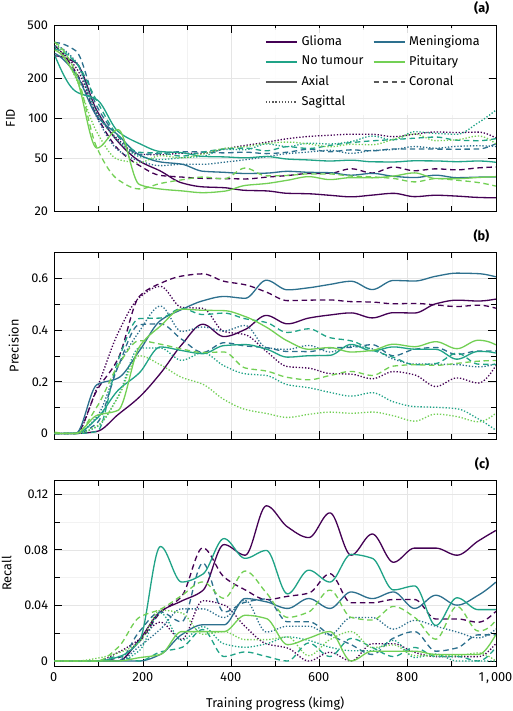}
  \caption{Training-time image-quality diagnostics for the twelve class-plane StyleGAN2-ADA generators. Each curve represents one tumour-class and anatomical-plane generator; colour indicates tumour class and line style indicates axial, coronal, or sagittal plane. Panel \textbf{(a)} shows FID across training progress; lower values indicate closer feature-distribution agreement with the corresponding real subset. Panel \textbf{(b)} shows precision, measuring whether generated samples lie near the real-image manifold. Panel \textbf{(c)} shows recall, measuring whether generated samples cover the diversity of the real-image manifold. The figure is an at-a-glance view of generator behaviour; the selected-checkpoint values used for image generation are listed numerically in Table~\ref{tab:gan-selected-metrics}. The key pattern is that FID and precision can look encouraging while recall remains low, which is why the paper treats downstream classifier utility as the decisive test rather than assuming image-quality metrics are sufficient.}
  \label{fig:gan-metrics}
  \figcaprule
\end{figure}

Across the twelve selected checkpoints, FID ranged from 25.36 to 54.18, KID from 0.00546 to 0.02794, precision from 0.3007 to 0.6178, and recall from 0.0097 to 0.0863. The FID and KID figures suggest several generators achieved reasonable feature-space alignment with their real subsets, but the recall numbers tell the more important story: even the best generator covered less than 9\% of the real manifold. With training sets as small as 310 images per generator, this is expected, and it is the main reason downstream augmentation gains cannot be read off favourable image-quality metrics alone (Fig.~\ref{fig:gan-metrics}).

The class-level downstream results reinforce that caution. The glioma generators had the highest mean recall across planes (0.065), yet glioma F1 did not show the largest augmentation gains in either deep classifier. The no-tumour generators had only moderate mean recall (0.040), and yet no-tumour F1 improved most consistently in the CNN (+2.58\% to +4.29\% across augmented conditions) and MobileViTV2 (+4.11\% to +4.67\%). Meningioma mean generator recall was lower still (0.038); meningioma F1 nonetheless improved for the deep classifiers, even as it remained the hardest class overall. The takeaway is straightforward: generator precision and recall diagnose feature-space quality, but they do not map one-to-one onto class-specific classifier gains.

\begin{table}[!tbp]
  \floatcaprule
\caption{Selected StyleGAN2-ADA checkpoints and generator quality metrics for the twelve tumour-class and anatomical-plane generators. The kimg column gives the snapshot used to generate synthetic images for downstream augmentation. FID and KID quantify real-versus-generated feature-distribution discrepancy; lower is closer. Precision estimates how often generated samples fall near the real-image manifold, and recall estimates how much of the real-image manifold the generated samples cover. The table separates visual fidelity from distributional coverage, an important distinction because downstream utility cannot be inferred from FID alone.}
\label{tab:gan-selected-metrics}
\centering
\normalfont\plotfigurefont\scriptsize
\begin{tabularx}{\linewidth}{Yrrrrr}
\toprule
Generator & kimg$^{\dagger}$ & FID & KID & Precision & Recall \\
\midrule
Glioma (Axial) & 960 & 25.36 & 0.00919 & 0.5128 & 0.0863 \\
Glioma (Coronal) & 336 & 35.27 & 0.00906 & 0.6178 & 0.0814 \\
Glioma (Sagittal) & 240 & 53.05 & 0.01971 & 0.5693 & 0.0279 \\
Meningioma (Axial) & 1,000 & 36.11 & 0.01016 & 0.6066 & 0.0567 \\
Meningioma (Coronal) & 240 & 54.18 & 0.01774 & 0.4243 & 0.0352 \\
Meningioma (Sagittal) & 240 & 43.93 & 0.01816 & 0.4931 & 0.0229 \\
No tumour (Axial) & 672 & 48.28 & 0.01572 & 0.3446 & 0.0767 \\
No tumour (Coronal) & 288 & 51.84 & 0.01512 & 0.4839 & 0.0097 \\
No tumour (Sagittal) & 240 & 48.90 & 0.01981 & 0.3306 & 0.0346 \\
Pituitary (Axial) & 288 & 28.40 & 0.00546 & 0.4811 & 0.0211 \\
Pituitary (Coronal) & 192 & 29.38 & 0.00849 & 0.3590 & 0.0314 \\
Pituitary (Sagittal) & 192 & 46.39 & 0.02794 & 0.3007 & 0.0134 \\
\bottomrule
\end{tabularx}
{\normalfont\plotfigurefont\scriptsize\raggedright $^{\dagger}$Selected snapshot in thousands of training images (kimg). All twelve runs were trained to the same 1,000-kimg budget; values below 1,000 indicate that an earlier snapshot was selected for image generation and metric reporting, not that the run was stopped early.\par}
\tabnoterule
\end{table}

Synthetic images were generated from the exponential-moving-average generator weights of the selected checkpoint for each tumour-plane model. Because each generator was trained on exactly one class-plane subset, generated images inherited their class and plane labels from the generator that produced them. Generation used truncation $\psi = 1.00$\footnote{Truncation $\psi=1$ leaves latent codes unmodified; the standard StyleGAN truncation trick, which interpolates sampled latents toward the distribution mean to trade diversity for visual quality, is not applied at this value \cite{karras2020stylegan2}.} and random noise mode, preserving diversity rather than intentionally trading diversity for visual sharpness. Latent vectors were generated deterministically from a base seed of 42 plus the candidate index, so that image generation was reproducible while still spanning the latent space.

\begin{figure}[!tbp]
  \floatcaprule
  \centering
  \includegraphics[width=0.79\linewidth]{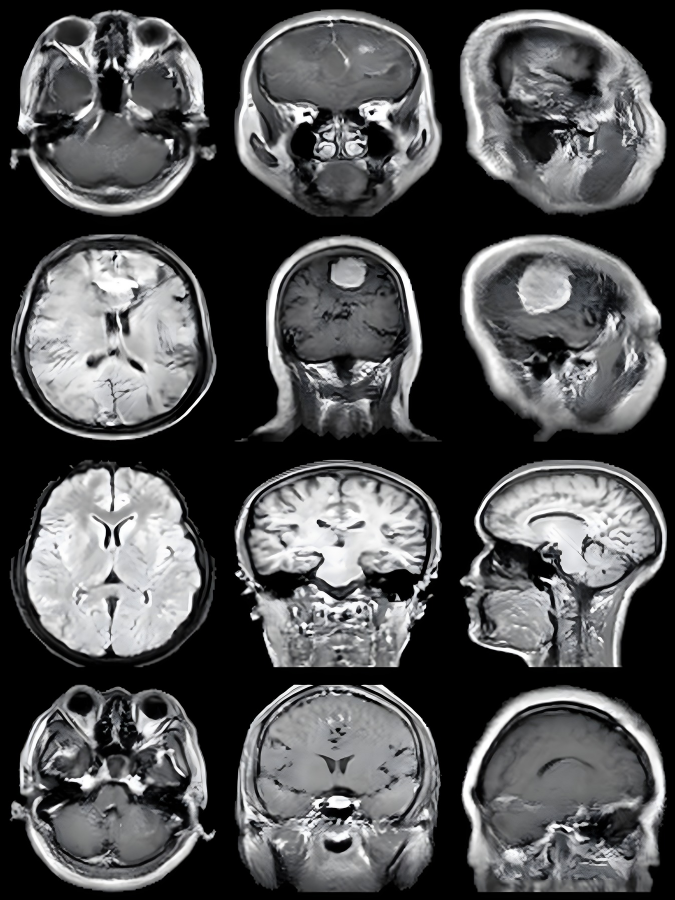}
  \caption{\textit{AI-generated images.} Representative synthetic MRI images generated by StyleGAN2-ADA from the selected class-plane checkpoints, after the same preprocessing pipeline applied to real BRISC images. The 4~$\times$~3 mosaic mirrors the experimental design: rows are tumour classes and columns are anatomical planes, so each cell shows an image sampled from one independent class-plane generator. The figure is qualitative and illustrative, not a diagnostic validation panel; it shows the visual form of the synthetic images entering the downstream augmentation experiments. Images are displayed here at a resolution higher than 128~$\times$~128 for illustrative purposes; 128~$\times$~128 was the working resolution used for all GAN training, generation, and downstream classification.}
  \label{fig:synthetic-mosaic}
  \figcaprule
\end{figure}

Candidate images were preprocessed immediately after generation using the same grayscale masking, intensity normalization, cropping, resizing, and remasking pipeline applied to real images (Fig.~\ref{fig:synthetic-mosaic}). Candidates were rejected before saving if their mean grayscale intensity was $\bar{x}<30$ or if their nonzero-pixel fraction was $f_{\mathrm{nz}}<0.08$ after preprocessing, because such images were effectively blank or structurally unusable. Perceptual-hash deduplication was then applied: a candidate was rejected if its 64-bit DCT-based perceptual hash (pHash, computed via \texttt{imagehash.phash} with default 8~$\times$~8 hash size) had Hamming distance $d_H \leq 6$ from any previously saved synthetic image or from any real train/test reference image in the same tumour-plane subset. This prevented the synthetic pool from being inflated by near-duplicates and reduced the risk of accidentally regenerating a training or held-out reference.

\subsection{Synthetic Filtering}

Synthetic filtering was implemented as a quality-control and diversity-selection stage, not as an assumed improvement. A synthetic pool multiplier of $\rho_{\mathrm{pool}}=2.5$ was used relative to the real training size so that, after rejection, enough candidates remained to construct both 1:1 and 1:2 filtered augmentation sets. Filtering was performed separately within each tumour-plane subset. For each subset, InceptionV3 pool3 features were computed for the real and candidate synthetic images. The real features were reduced with unwhitened PCA using $k_{\mathrm{PCA}}\leq \min(200,n_{\mathrm{real}}-1)$ principal components; because every class-plane subset held at least 310 real images, the effective value was $k_{\mathrm{PCA}}=200$ for all twelve filters. A Ledoit-Wolf covariance model \cite{ledoitwolf2004}\footnote{The Ledoit-Wolf estimator is a regularized covariance matrix estimator that shrinks the sample covariance toward a scaled identity target, improving conditioning in the high-dimensional, small-sample setting typical of InceptionV3 feature spaces.} was then fitted to the PCA-reduced real features.

Synthetic candidates were scored by squared Mahalanobis distance from the PCA-reduced real subset distribution:
\begin{equation}
  D^2_i = (\mathbf{x}_i - \boldsymbol{\mu})^\top \boldsymbol{\Sigma}^{-1} (\mathbf{x}_i - \boldsymbol{\mu}),
\end{equation}
where $\boldsymbol{\mu}$ is the real feature mean and $\boldsymbol{\Sigma}$ is the Ledoit-Wolf covariance estimate. Rather than using a theoretical chi-square cutoff, the threshold was calibrated empirically as $q_{0.975}$, the 97.5\textsuperscript{th} percentile of the real images' own squared Mahalanobis distances under the same model. A synthetic image was rejected only when $D_i^2>q_{0.975}$, i.e. when it was more feature-atypical than the most atypical 2.5\% of real images. This made the filter self-calibrating per subset and avoided assuming multivariate normality. Among candidates that passed the Mahalanobis gate, deterministic farthest-point (greedy $k$-centre) sampling selected a diverse subset: the algorithm seeded with the surviving synthetic image whose PCA-reduced feature vector was closest to the survivor centroid, then iteratively added the candidate whose minimum distance to any already-selected image was greatest. This prevents the filtered set from collapsing into many visually similar high-density samples.

This process produced nested filtered sets: the filtered 1:1 subset is contained within the filtered 1:2 subset. From a pool of 12,239 synthetic images,\footnote{The nominal pool size is $2.5 \times 4{,}896 = 12{,}240$; the one-image shortfall arises because per-generator quotas are computed by rounding the class-plane training counts individually, producing integer totals that sum to 12,239 rather than 12,240.} the $q_{0.975}$ Mahalanobis gate rejected 992 candidates (8.11\%). After farthest-point selection, the filtered 1:1 condition retained 4,896 images and the filtered 1:2 condition retained 9,792. Threshold audits were also generated for $q_{0.95}$ and $q_{0.99}$, which rejected 1,347 (11.01\%) and 717 (5.86\%) candidates respectively; however, downstream classifiers were run only for $q_{0.975}$. Because filtering can remove both harmful artifacts and potentially useful variation, its value was tested experimentally rather than presumed, and cutoff sensitivity remains a follow-up experiment rather than a claim of this paper.

\begin{figure}[!tbp]
  \floatcaprule
  \centering
  \includegraphics[width=\linewidth]{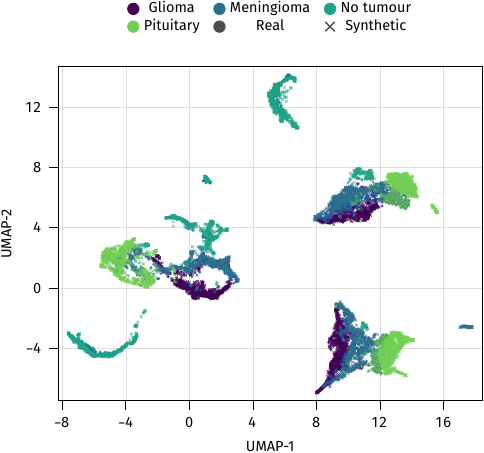}
  \caption{UMAP projection of InceptionV3 pool3 features for all real training images and all synthetic pool images used in the filtering stage. Filled circles are real images, cross marks are synthetic candidates, and colour indicates tumour class. The projection was fitted jointly to real and synthetic features using two output dimensions, Euclidean distance, 15 nearest neighbours, minimum embedding distance 0.1, and random seed 42; distances consequently reflect a shared feature space rather than separate within-source embeddings. Broad overlap between real and synthetic points of the same class indicates that the generators captured class-relevant structure, while partial non-overlap indicates incomplete coverage. The diffuse \textit{meningioma} region is especially informative: it agrees with the low recall values in Table~\ref{tab:gan-selected-metrics} and motivates the feature-space filtering applied before constructing the filtered augmentation sets.}
  \label{fig:umap-class}
  \figcaprule
\end{figure}

Figure~\ref{fig:umap-class} visualizes the joint InceptionV3 feature space of real and synthetic training images via UMAP \cite{mcinnes2018umap}. Real and synthetic images from the same class-plane generator occupy broadly overlapping regions, confirming that synthetic data was not pushed to a separate manifold. \textit{Meningioma} images occupy a more diffuse cluster than \textit{glioma} and \textit{pituitary}, consistent with the higher per-class variability in meningioma anatomy. The partial rather than perfect overlap motivates the Mahalanobis-distance filter, which retains synthetic images whose features are consistent with the real class distribution and discards outlier samples.

\subsection{Downstream Classifiers}

Three classifier families were chosen to probe whether augmentation effects depended on architecture: a fixed-feature classical model, a convolutional network trained from scratch on raw images, and a pretrained hybrid transformer.

The fixed-feature baseline was a random forest \cite{breiman2001rf} operating on $d=2{,}048$-dimensional InceptionV3 \cite{szegedy2016inception} pool3 descriptors. Preprocessed images were resized to 299~$\times$~299 before feature extraction; features were passed to the forest without additional normalization or dimensionality reduction, since decision-tree ensembles are invariant to monotone transforms. The forest used $T=500$ trees, $m_{\mathrm{try}}=\log_2(d)$ candidate features per split, minimum leaf size $n_{\min}=2$, bootstrap sampling, out-of-bag scoring, warm-start growth, and $N_{\mathrm{seed}}=10$ independent seeds. This model does not test raw-pixel representation learning: if it benefits from synthetic augmentation, the benefit must come through the geometry of a fixed InceptionV3 feature space. Moreover, ImageNet-pretrained descriptors transfer partially but imperfectly to MRI \cite{tajbakhsh2016finetuning}, which makes the RF a deliberately conservative baseline.

\begin{figure}[!tbp]
  \floatcaprule
  \centering
  \includegraphics[width=0.66\linewidth]{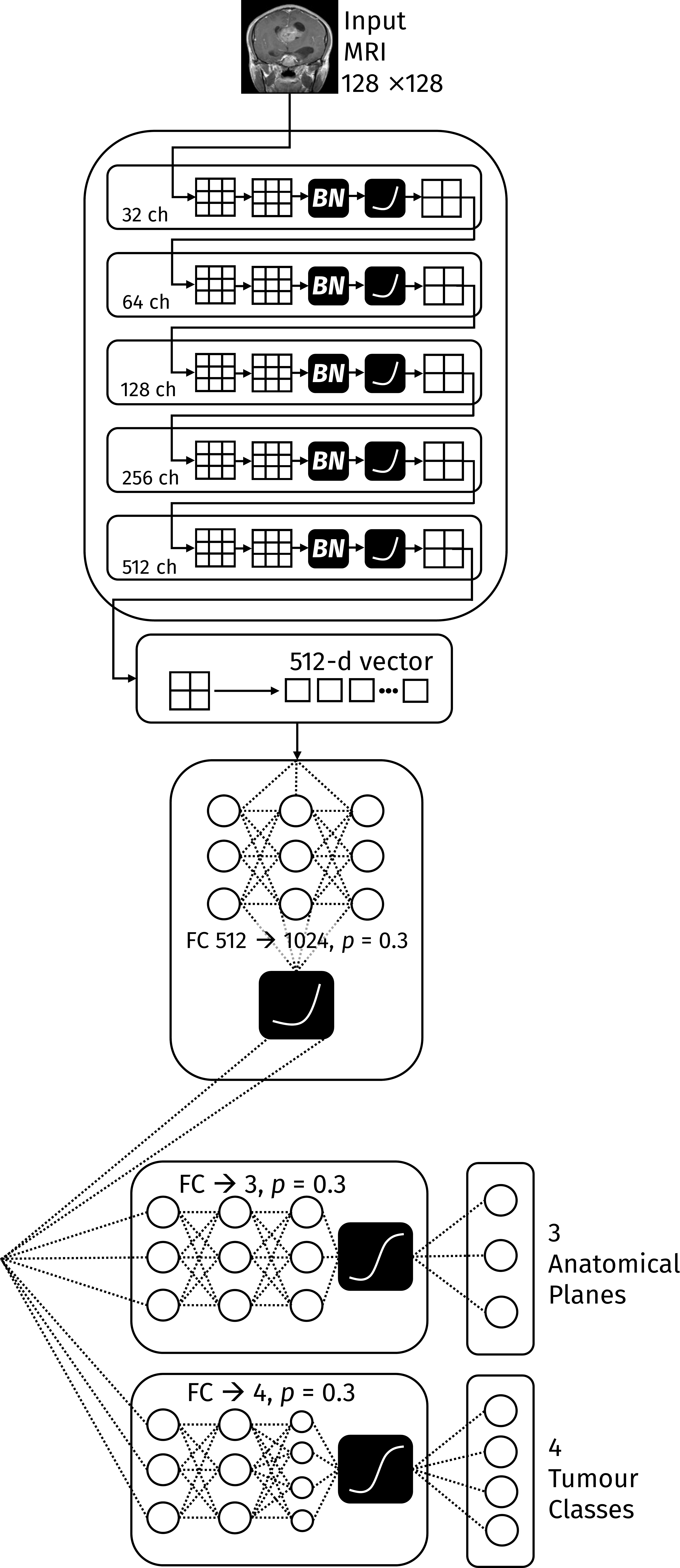}
  \caption{Compact two-headed CNN used as the end-to-end convolutional classifier. A 128~$\times$~128 preprocessed MRI slice enters a shared five-block convolutional backbone; each block applies two 3~$\times$~3 convolutions, batch normalization, ReLU activation, and 2~$\times$~2 max pooling. Global adaptive average pooling collapses the final feature maps into a 512-dimensional representation before a shared fully connected layer maps 512~$\to$~1024 with dropout probability $p_{\mathrm{drop}}=0.3$ and ReLU. The shared representation then splits into two supervised heads. The tumour head applies dropout, FC~$\to$~4, and softmax to produce four tumour-class probabilities; the anatomical-plane head applies dropout, FC~$\to$~3, and softmax to produce three plane probabilities. The output boxes are drawn outside the heads to emphasize that they are final predictions rather than additional hidden layers.}
  \label{fig:cnn-architecture}
  \figcaprule
\end{figure}

The second classifier was a compact two-headed CNN trained end-to-end on 128~$\times$~128 preprocessed images (Fig.~\ref{fig:cnn-architecture}). The baseline design was adapted from the brain-tumour MRI CNN of Ali and Behzad \cite{ali2025dcgancnn}, which paired repeated Conv2D--batch-normalisation--pooling blocks with global average pooling and a 1024-unit dense layer for binary tumour classification. That compact convolutional template was retained but the architecture was made task-specific: the binary sigmoid output was replaced by two supervised softmax heads, one for four-way tumour classification and one for three-way anatomical-plane classification. The backbone comprised five convolutional blocks with channel vector $\mathbf{c}=(32,64,128,256,512)$, each with two 3~$\times$~3 convolutions (padding~1), batch normalisation, ReLU, and 2~$\times$~2 max pooling. After global adaptive average pooling to a 512-dimensional representation, a shared fully connected layer (512~$\to$~1024, dropout probability $p_{\mathrm{drop}}=0.3$, ReLU) fed the two independent classification heads, each with its own dropout layer using $p_{\mathrm{drop}}=0.3$. Adam was used with learning rate $\eta = 10^{-3}$, moment coefficients $(\beta_1,\beta_2)=(0.9,0.999)$, numerical constant $\epsilon_{\mathrm{Adam}}=10^{-8}$, weight decay $\lambda = 10^{-3}$, batch size $B=256$, and label smoothing $\epsilon_{\mathrm{LS}}=0.1$. Class and plane weights were computed from real-only training rows as
\begin{equation}
  w_c = \min\!\left(\bar{n}/n_c,\; 5.0\right),
\end{equation}
where $\bar{n}$ is the mean class count and $n_c$ is the real-sample count for class $c$, anchoring the loss prior to the real distribution regardless of synthetic ratio. Mixup was not used. Validation-loss-based checkpointing ran over a maximum of $E_{\max}=200$ epochs with early-stopping patience $P_{\mathrm{ES}}=30$; if validation loss plateaued for $P_{\eta}=10$ epochs, the scheduler updated $\eta \leftarrow \eta/2$.

Both end-to-end models used an uncertainty-weighted multitask loss with learned log-variance terms \cite{kendall2018multitask}. For tumour loss $L_t$ and plane loss $L_p$, the optimized objective was
\begin{equation}
  \mathcal{L} = e^{-s_t}L_t + s_t + e^{-s_p}L_p + s_p,
\end{equation}
where $s_t$ and $s_p$ are learned scalar log variances. This lets the model learn the relative weighting of tumour and plane supervision rather than fixing it manually. Because of the additive log-variance terms, the reported training and validation losses can become negative; they should not be read as ordinary cross-entropy values.

\begin{figure}[!tbp]
  \floatcaprule
  \centering
  \includegraphics[width=\linewidth,height=0.65\textheight,keepaspectratio]{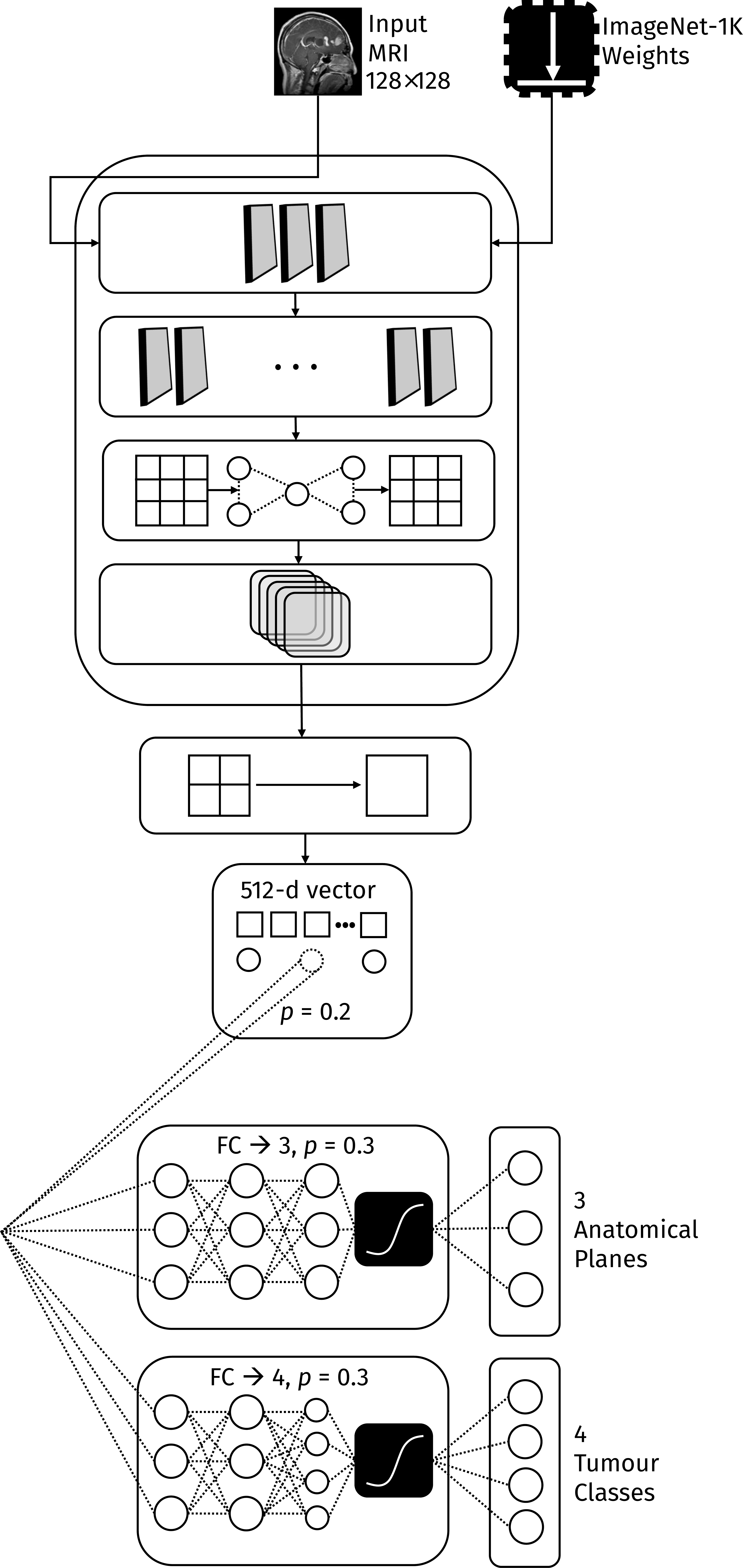}
  \caption{MobileViTV2 control architecture used as the pretrained hybrid convolutional-transformer classifier. A 128~$\times$~128 preprocessed MRI slice enters an ImageNet-initialised MobileViTV2-100 CVNets backbone: the stacked modules depict the convolutional stem, mobile convolution blocks, separable self-attention over token grids, and the final feature-map stack. Adaptive average pooling reduces the shared feature maps to a 512-dimensional representation, followed by dropout with $p_{\mathrm{drop}}=0.2$. The representation then branches into two supervised heads: FC~$\to$~4 plus softmax for tumour classification and FC~$\to$~3 plus softmax for anatomical-plane classification. Unlike the CNN, MobileViTV2 was trained under a compute-matched step budget across baseline and augmented conditions, making it a higher-capacity test of whether synthetic data can improve representation learning without simply giving augmented runs more optimizer updates.}
  \label{fig:mobilevit-architecture}
  \figcaprule
\end{figure}

MobileViTV2 served as the higher-capacity hybrid convolutional-transformer control \cite{mehta2022mobilevitv2} (Fig.~\ref{fig:mobilevit-architecture}). The backbone loaded in the experiment was the MobileViTV2-100 CVNets ImageNet-1k variant with 512 output features; the exact timm identifier is retained in the saved configuration artifacts. A fallback chain was kept in the implementation for portability, but no fallback variant was used in the reported runs. The model was adapted for the two-task setup by replacing the original classifier with two independent heads after adaptive average pooling and dropout probability $p_{\mathrm{drop}}=0.2$. It used AdamW \cite{loshchilov2019adamw} with learning rate $\eta = 5\times10^{-4}$, moment coefficients $(\beta_1,\beta_2)=(0.9,0.999)$, numerical constant $\epsilon_{\mathrm{AdamW}}=10^{-8}$, weight decay $\lambda = 10^{-4}$, batch size $B=128$, label smoothing $\epsilon_{\mathrm{LS}}=0.001$, cosine learning-rate scheduling \cite{loshchilov2017sgdr} with warmup length $E_{\mathrm{warm}}=5$ epochs, L2-norm gradient clipping $\|\nabla\|_2 \leq 1.0$, Mixup augmentation \cite{zhang2018mixup} with concentration $\alpha_{\mathrm{mix}}=0.2$ and mixing coefficient $m\sim\mathcal{B}(\alpha_{\mathrm{mix}},\alpha_{\mathrm{mix}})$, and automatic mixed precision by default. Unlike the epoch-based CNN, MobileViTV2 used a compute-matched step budget across baseline and augmentation conditions. The default maximum step count was derived from $E=55$ real-only epochs. With approximately 3,917 real training rows after the validation split and $B=128$, one real-only budget epoch contained $\lceil 3917/128\rceil=31$ optimizer steps, giving $55\times31=1{,}705$ maximum steps. In the completed runs, the real-only selected checkpoints occurred at 53.2~\textpm{}~2.6 real-data epochs, so the 55-epoch ceiling functioned as a near-full real-only training budget rather than an augmented-arm advantage. Validation occurred every $v=80$ optimizer steps, with early-stopping patience $P_{\mathrm{ES}}=8$ validation evaluations and minimum improvement $\delta_{\min}=0.005$. For 1:1 augmentation, each batch mixed $B_{\mathrm{real}}=64$ real and $B_{\mathrm{synth}}=64$ synthetic images; for 1:2 augmentation, each batch mixed $B_{\mathrm{real}}=43$ real and $B_{\mathrm{synth}}=85$ synthetic images.\footnote{With a total batch size of $B=128$ and a target 1:2 ratio, the integer split is $\lceil 128/3 \rceil = 43$ real and $128-43=85$ synthetic, yielding an achieved ratio of $1{:}1.98$.} This prevented augmented runs from receiving a larger effective training budget solely because their datasets were larger.

Each classifier family was evaluated under five conditions (real-only baseline, unfiltered 1:1, unfiltered 1:2, filtered 1:1, and filtered 1:2) against the same held-out BRISC test set and across ten independent seeds, $\{0,\ldots,9\}$. This design tests the stated question: whether generated images provide utility beyond training on the real cohort alone. It does not test whether GAN-based supplementation outperforms conventional transform-based augmentation, which is a separate comparison requiring its own controlled design. For the end-to-end models, each seed controlled the real-only train/validation split, model initialization, mini-batch order, Mixup pairing where enabled, and the NumPy, Python, and PyTorch/CUDA random states. For the RF, the seed controlled the random forest bootstrap and split sampling through the model's random state.

Validation data were always drawn from real images only. For each CNN and MobileViTV2 seed, 20\% of the real training manifest was reserved as a validation set using an 80/20 split stratified by the combined tumour-plane label; if a rare stratum made stratification impossible, the code fell back to an unstratified 80/20 split for that seed. Synthetic images were never included in validation and the held-out BRISC test set was never used for checkpoint selection. This makes the selected checkpoints comparable across real-only and augmented conditions: augmentation could influence training, but the validation signal remained real-only.

The two end-to-end classifiers mixed synthetic images differently because their training budgets were defined differently. The CNN used an epoch-based schedule over a shuffled appended manifest: after the real validation rows were removed, the remaining real training rows and the required synthetic rows were concatenated and sampled by the ordinary shuffled data loader. CNN mini-batches were not forced to contain a fixed real/synthetic quota, although the epoch-level manifest ratio matched the intended augmentation condition. MobileViTV2 used explicit per-step quotas so that the optimizer-step budget remained compute-matched: 1:1 batches used 64 real and 64 synthetic images, and 1:2 batches used 43 real and 85 synthetic.

Tumour classification accuracy was the primary endpoint; macro F1, weighted F1, per-class F1, and plane accuracy were secondary. Statistical comparisons used two-tailed paired case-level permutation tests with 5,000 resamples. For a condition $A$ and its baseline $B$, predictions were stored as matrices over seed $s$ and held-out test image $i$. The observed statistic was the difference between the mean metric over seeds for $A$ and $B$. Under the null hypothesis that the two conditions are exchangeable for each held-out image, the labels $A$ and $B$ were independently swapped with probability 0.5 at the image level across all seeds, producing the permutation distribution. Percentile bootstrap confidence intervals were computed by resampling held-out test images with replacement. Holm correction was then applied across all four augmentation conditions crossed with all primary and secondary metrics within each classifier family. Seed-level summaries and distribution plots are reported as stability diagnostics, but the ten seed means were not used as the sole exchangeability units; doing so would discard the paired held-out case structure and answer a lower-resolution question.

The saved training-history JSONL files were also analysed as a secondary efficiency endpoint. For the CNN, which used epoch-based training, convergence efficiency was summarized by the epoch of the validation-loss-selected checkpoint. For MobileViTV2, optimizer steps were deliberately compute-matched across baseline and augmented conditions, so step count is not a fair measure of faster convergence. Efficiency was instead measured as the number of real-data epochs seen by the time the selected checkpoint was reached: baseline batches contained 128 real images, 1:1 batches contained 64 real and 64 synthetic, and 1:2 batches contained 43 real and 85 synthetic. The relevant question for MobileViTV2 is instead whether a comparable or better selected checkpoint was reached with fewer passes over the scarce real data. Paired seed-level exact sign-flip tests compared baseline against each augmentation condition for these secondary efficiency measures. Seed stability was summarized as the standard deviation of held-out tumour accuracy across the ten seeds.

\subsection*{AI Writing Assistance}

Manuscript prose was drafted with the assistance of a large language model (LLM). The author prepared a detailed section-by-section outline specifying the points, arguments, and structure to be addressed; the LLM generated prose from those outlines. All generated text was reviewed, corrected, and approved by the author, who takes full responsibility for the accuracy and integrity of the final manuscript. No LLM is listed as an author. The LLM was not used to perform data collection, model training, statistical analysis, or primary figure generation.

\section{Results}

\subsection{StyleGAN2-ADA Behaviour and Synthetic Image Quality}

The selected StyleGAN2-ADA checkpoints produced coherent 128~$\times$~128 brain MRI-like images across all twelve tumour-plane combinations. The qualitative mosaics and selected-checkpoint metrics support the only claim the downstream experiment requires: the generated images were structured enough to test whether they add classifier utility.

The generator metrics expose a tension that recurs throughout the paper. Several FID values look encouraging in isolation, but precision and recall together tell a more nuanced story: the generators could produce plausible samples without covering much of the real distribution. That limited coverage---a direct consequence of training on only a few hundred images per generator---is the proximate reason downstream augmentation did not universally help.

\subsection{VLM Blind Test: Real vs.\ Synthetic Discrimination}

As an independent realism probe, GPT-5.5 was queried via the OpenAI Responses API on a balanced set of 2,016 preprocessed images (1,008 real and 1,008 synthetic) drawn from the augmented 1:2 training condition. Images were sampled with stratified balancing across all source$\times$class$\times$plane combinations (84 images per stratum, 24 strata). The final reported API batch was prepared, submitted, and parsed on May 7, 2026; the saved request metadata records the requested model string \texttt{gpt-5.5}, but the public endpoint did not expose an immutable model snapshot identifier in the saved response artifacts, so reproducibility is tied to the GPT-5.5 endpoint as accessed on that date. The final reported batch used a cost-controlled grouped prompt with \texttt{reasoning.effort="low"}, low text verbosity, and four images per request; each image was embedded as a base64 PNG data URL with \texttt{detail="original"}. Because the images themselves were 128~$\times$~128 preprocessed PNGs, original-detail submission preserved the available image resolution but did not restore information removed by preprocessing or resizing.

The exact system instruction was: ``You are a blinded visual-realism judge for a research-only audit of preprocessed 128x128 brain-MRI-like images. Some images are real and some are synthetic. Do not provide medical advice. Return only the requested JSON. Do not include explanations.'' The user instruction required one compact result per labelled image, stated that filenames and IDs were blinded random identifiers, and included the guardrail: ``Small, cropped, grayscale/RGB-encoded, normalized, or preprocessed appearance applies to both real and synthetic images; do not use that alone.'' Structured output was enforced by a strict JSON schema with one array entry per image. Each entry contained blinded identifiers, an image-usability flag, tumour and plane predictions with confidence scores, a binary source prediction with confidence, a 1--5 realism score, and artifact categories. Tumour class, plane, and source were restricted to the predefined labels used elsewhere in the experiment. The full schema is retained with the batch artifacts; in the manuscript, the important point is that the schema eliminated ambiguous prose responses and made the VLM probe reproducible as a structured batch audit.

The tumour-type and anatomical-plane predictions served as gate questions, confirming that the model had engaged with the image content before committing to the discrimination answer. Only images on which GPT-5.5 correctly identified both the tumour type and the anatomical plane were carried forward to the real-versus-synthetic analysis; images on which either gate was answered incorrectly were excluded, since the source prediction in those cases could not be trusted to reflect genuine visual reasoning.

Gate performance was 88.74\% for anatomical-plane identification and 52.08\% for tumour-type identification; the latter reflects the difficulty of a four-class problem without task-specific fine-tuning. Among the 899 images that passed both gates, GPT-5.5 achieved 57.73\% accuracy on the real-versus-synthetic discrimination task (519/899; Wilson 95\% CI: 54.48--60.92\%; two-sided binomial test against 50\%, $p<0.00001$). This is statistically above chance because the gated sample is large, but the effect size is modest: the synthetic images retained enough of the visual character of real MRI that a general-purpose vision-language model, confirmed to be attending to image content, still misclassified the source on 42.27\% of gated cases. GPT-5.5 was not asked to judge diagnostic quality or clinical plausibility; it was asked a binary discrimination question under controlled, low-output conditions. The figure is best interpreted as evidence of surface-level realism at the preprocessed working resolution, not proof of clinical indistinguishability.

The gating design also introduces a selection bias that should be read conservatively. Only 899 of 2,016 images (44.59\%) passed both content gates, so the source-discrimination analysis is conditioned on images whose tumour class and plane were visually legible to the model. Those images are likely clearer and less ambiguous than the excluded portion of the audit set, so the 57.73\% figure should be interpreted as a realism probe on the model-legible subset rather than a universal estimate over all generated images.

Per-class gated source-discrimination accuracy was 65.28\% for \textit{glioma}, 57.62\% for \textit{meningioma}, 56.44\% for \textit{no-tumour}, and 53.51\% for \textit{pituitary}. The class ordering is not interpreted as a generator-quality ranking; the probe was not designed for that purpose and the per-class gated sample sizes differ.

The VLM result complements the downstream classifier experiment. A VLM that easily separated real from synthetic would imply that the synthetic images carry source-identifying artifacts a downstream model could exploit as shortcuts. The near-chance gated figure does not rule out such artifacts, but it shows they are not conspicuous enough to be trivially detected by a general-purpose vision model attending to the images.

The next three subsections should be read accordingly as an architecture-dependent utility test rather than a single augmentation verdict. The same synthetic pool can be visually plausible and yet harmful to a fixed-feature classifier, cautiously useful to a compact CNN, and measurably useful to a pretrained hybrid model. That pattern is interpreted through a bias--variance lens in Section~\ref{sec:discussion-main}.

\subsection{Random Forest}

\begin{figure}[!tbp]
  \floatcaprule
  \centering
  \includegraphics[width=\linewidth]{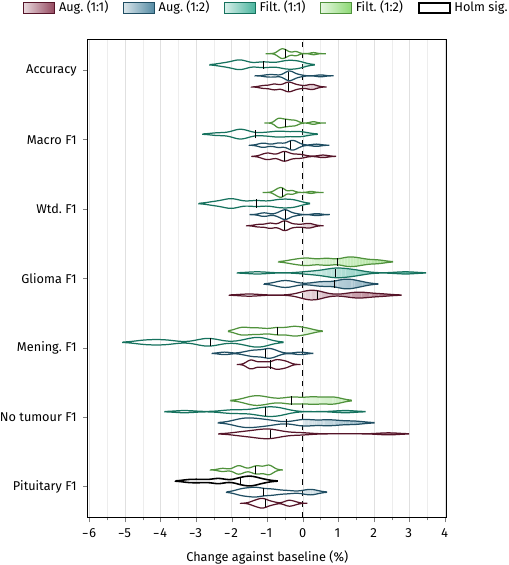}
  \caption{Paired random-forest changes relative to the real-only baseline. Each row is a tumour-class or aggregate metric, and each horizontal violin shows the seed-level change for one augmentation condition: unfiltered 1:1, unfiltered 1:2, filtered 1:1, or filtered 1:2. The vertical dashed zero line marks no change; values to the right favour augmentation and values to the left favour the real-only baseline. Negative portions of a violin are left unfilled, while positive portions are filled with that condition's colour using a zero-anchored gradient: the fill begins very lightly at 0\% and intensifies as the positive change grows. Black-outlined violins indicate comparisons that remained significant after Holm correction. The RF did not benefit from synthetic augmentation, and the filtered 1:1 condition worsened several class-level outcomes, showing that feature-space filtering can still be mismatched to a downstream model's decision surface.}
  \label{fig:rf-deltas}
  \figcaprule
\end{figure}

The random forest achieved a real-only baseline accuracy of 80.69\% \textpm{} 0.39\% (Table~\ref{tab:main-results}). Neither unfiltered nor filtered synthetic augmentation improved this primary endpoint: mean accuracy was 80.30\% for unfiltered 1:1, 80.32\% for unfiltered 1:2, 79.59\% for filtered 1:1, and 80.29\% for filtered 1:2 (Table~\ref{tab:main-results}). None of the primary accuracy changes remained significant after Holm correction (Table~\ref{tab:primary-paired}). The strongest negative secondary result was filtered 1:1 \textit{pituitary} F1, which decreased by 1.98\% absolute and was the only Holm-corrected significant comparison in the RF family (Fig.~\ref{fig:rf-deltas}).

\begin{figure}[!tbp]
  \floatcaprule
  \centering
  \includegraphics[width=\linewidth]{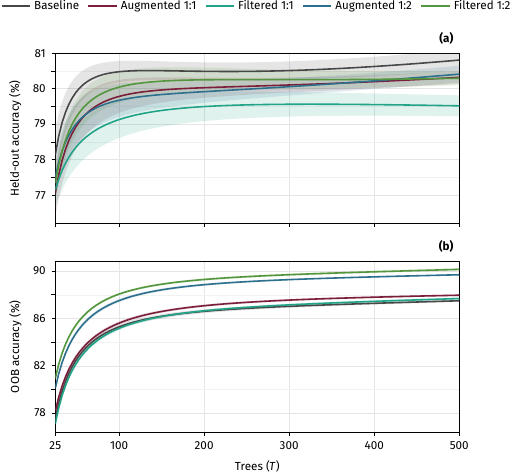}
  \caption{Random-forest behaviour as the number of trees increased from 25 to 500. Each panel shows all five training conditions (real-only baseline, unfiltered 1:1, unfiltered 1:2, filtered 1:1, and filtered 1:2), with shaded bands showing estimated 95\% CI across ten seeds. Panel \textbf{(a)} shows held-out tumour-classification accuracy on the fixed BRISC test set; panel \textbf{(b)} shows out-of-bag (OOB) accuracy, computed internally from bootstrap-excluded samples during random-forest training. The OOB bands are present but visually narrow because the seed-wise OOB estimates converge tightly by the final tree counts. The panels answer different questions: held-out accuracy is the reported endpoint, whereas OOB accuracy is a training-internal validation-like diagnostic. Both curves saturate early, indicating that the negative RF augmentation result was not caused by too few trees.}
  \label{fig:rf-tree-growth}
  \figcaprule
\end{figure}

\begin{figure}[!tbp]
  \floatcaprule
  \centering
  \includegraphics[width=\linewidth]{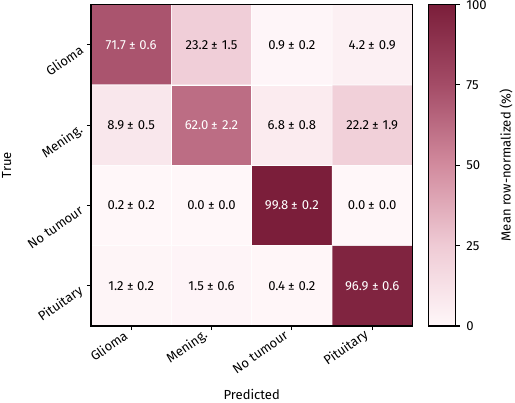}
  \caption{Seed-averaged tumour confusion matrix for the random forest under the real-only baseline condition. Rows are true tumour classes and columns are predicted tumour classes; each cell is the row-normalized percentage of held-out test images assigned to that predicted class, reported as $\bar{x}$ \textpm{} half 95\% CI across ten seeds. The diagonal gives class recall. The RF was highly reliable for \textit{no tumour} (99.8\% recall) but weak for \textit{meningioma} (62.0\% recall), which was frequently confused with \textit{glioma} (23.2\%) and \textit{pituitary} (22.2\%); this identifies the main class-level limitation of the fixed-feature baseline.}
  \label{fig:rf-confusion}
  \figcaprule
\end{figure}

The random forest tree-growth plot adds a useful diagnostic (Fig.~\ref{fig:rf-tree-growth}). Held-out accuracy stabilized quickly as trees were added, and out-of-bag (OOB) accuracy increased smoothly. OOB accuracy was not used to select the final held-out result; it is reported only as an internal, validation-like estimate generated from bootstrap-excluded samples. The baseline tumour confusion matrix (Fig.~\ref{fig:rf-confusion}) shows that \textit{meningioma} was the hardest class for the RF, with 62.0\% recall and 23.2\% of cases misclassified as \textit{glioma} and 22.2\% as \textit{pituitary}. \textit{No-tumour} cases were classified near-perfectly (99.8\%).

The RF outcome is best read as a boundary case for synthetic augmentation rather than a failed run. The tree-growth and OOB curves show that the model had reached stable capacity by the final forest size, so the negative result is not a late tree-count artifact. The persistent negative or near-zero deltas instead indicate that synthetic samples did not sharpen the RF decision regions and may have blurred class boundaries for the harder classes, particularly \textit{meningioma} and \textit{pituitary}. The contrast also helps frame the later CNN and MobileViTV2 gains as architecture-specific effects rather than evidence that the generated images are intrinsically useful for every classifier.

% Float placement is intentionally left flexible here so LaTeX can pack the results section tightly.

\subsection{CNN}

The compact CNN achieved a real-only tumour accuracy of 93.92\% \textpm{} 0.45\% (Table~\ref{tab:main-results}). Mean tumour accuracy rose under every synthetic condition: 94.92\% for unfiltered 1:1, 94.99\% for unfiltered 1:2, 94.70\% for filtered 1:1, and 94.67\% for filtered 1:2. The mean gains are consistent, but none of the primary tumour-accuracy comparisons remained significant after Holm correction (Table~\ref{tab:primary-paired}). The CNN consequently occupies an intermediate position: consistent mean improvement, without enough corrected statistical evidence to claim a robust primary-accuracy benefit. The training histories, however, tell a different story than the final accuracy table---all four augmented CNN conditions selected their best-validation-loss checkpoints much earlier than baseline (Table~\ref{tab:efficiency-stability}).

\begin{figure}[!b]
  \floatcaprule
  \centering
  \includegraphics[width=\linewidth]{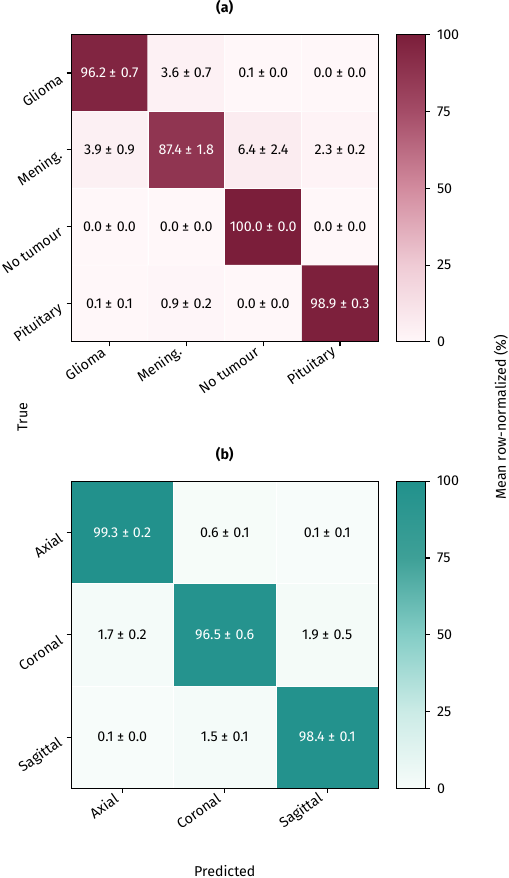}
  \caption{Seed-averaged confusion matrices for the compact two-headed CNN under the real-only baseline condition. Panel \textbf{(a)} is the tumour-class head: rows are true tumour classes and columns are predicted tumour classes. Panel \textbf{(b)} is the anatomical-plane head: rows are true planes and columns are predicted planes. Cell values are row-normalized percentages, reported as $\bar{x}$ \textpm{} half 95\% CI across ten seeds, so diagonal cells are per-class or per-plane recall. The CNN substantially improved \textit{meningioma} recall relative to the RF (87.4\% vs.\ 62.0\%) and classified anatomical plane near perfectly.}
  \vspace{0.40\baselineskip}
  \label{fig:cnn-confusion}
  \figcaprule
\end{figure}

\begin{figure}[!tbp]
  \floatcaprule
  \centering
  \includegraphics[width=\linewidth]{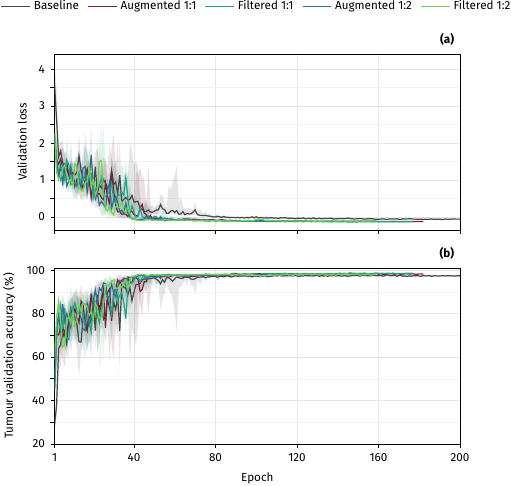}
  \caption{Validation behaviour of the compact two-headed CNN across the five training conditions. Panel \textbf{(a)} shows the multitask validation loss used for checkpoint selection; panel \textbf{(b)} shows tumour-head validation accuracy over the same epochs. Solid curves are seed medians and shaded bands are interquartile ranges (25\textsuperscript{th}--75\textsuperscript{th} percentile) across ten seeds, covering full training histories rather than just selected checkpoints. The loss is not ordinary cross-entropy because the CNN used an uncertainty-weighted tumour-plus-plane objective with learned log-variance terms, so negative validation loss values are expected. The dominant pattern is not a large separation in final validation accuracy, but a shift in timing: all synthetic conditions reached their selected validation-loss checkpoint much earlier than the real-only baseline (Table~\ref{tab:efficiency-stability}).}
  \label{fig:cnn-curves}
  \figcaprule
\end{figure}

The CNN learning curves show validation accuracy rising quickly and staying high, while validation loss fluctuates under the uncertainty-weighted multitask objective (Fig.~\ref{fig:cnn-curves}). Because the loss includes learned task-uncertainty terms, negative values are expected and do not indicate an implementation error. The baseline confusion matrices (Fig.~\ref{fig:cnn-confusion}) show the CNN separating all four tumour classes substantially better than the RF, with \textit{meningioma} recall at 87.4\% and anatomical plane near-perfect.

\subsection{MobileViTV2}
\label{sec:results-mvitv2}

MobileViTV2 achieved a real-only tumour accuracy of 94.18\% \textpm{} 0.69\% and plane accuracy of 99.16\% \textpm{} 0.12\% (Table~\ref{tab:main-results}). Synthetic augmentation improved mean tumour accuracy in every condition. The largest primary gain came from filtered 1:1 augmentation: 94.98\% mean accuracy, paired case-level delta of 1.02\% absolute (95\% CI: 0.54--1.54\%), Holm-corrected $p = 0.0104$ (Table~\ref{tab:primary-paired}). Unfiltered 1:1 and unfiltered 1:2 also remained significant for tumour accuracy after Holm correction; filtered 1:2 improved mean accuracy, but the change did not survive correction. MobileViTV2 also showed the clearest seed-stability gain: all four augmented conditions reduced the seed-level standard deviation of held-out tumour accuracy while preserving or improving the mean.

\begin{figure}[!tbp]
  \floatcaprule
  \centering
  \includegraphics[width=\linewidth]{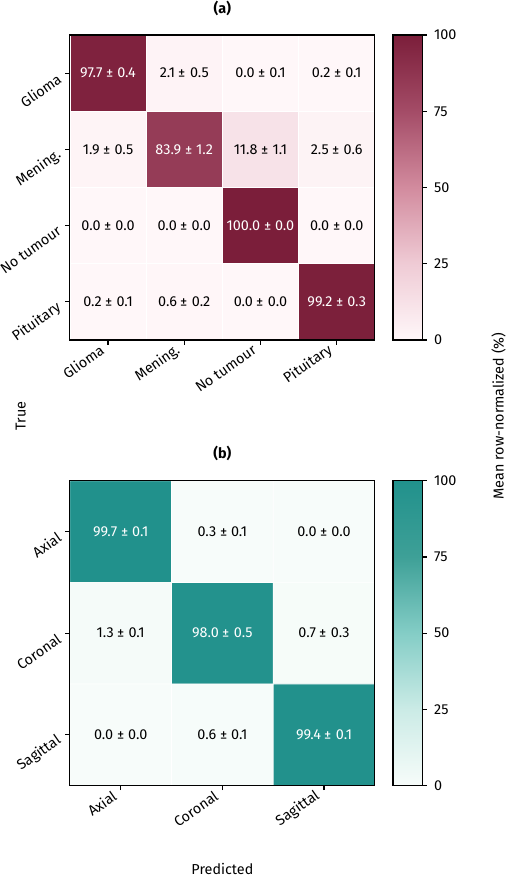}
  \caption{Seed-averaged confusion matrices for MobileViTV2 under the real-only baseline condition. Panel \textbf{(a)} reports the tumour-class head, with true tumour classes in rows and predicted tumour classes in columns. Panel \textbf{(b)} reports the anatomical-plane head, with true planes in rows and predicted planes in columns. Cell values are row-normalized percentages, reported as $\bar{x}$ \textpm{} half 95\% CI across ten seeds; diagonal cells are recalls. MobileViTV2 achieved high recall across most tumour classes, but \textit{meningioma} remained the weakest class (83.9\% recall), primarily through confusion with \textit{no tumour} (11.8\%).}
  \label{fig:mobilevit-confusion}
  \figcaprule
\end{figure}

\begin{figure}[!tbp]
  \floatcaprule
  \centering
  \includegraphics[width=\linewidth]{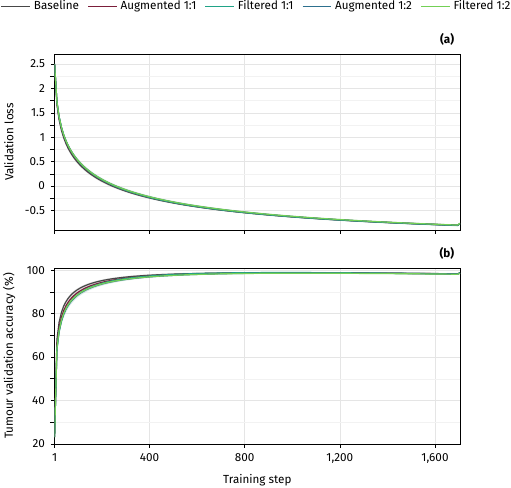}
  \caption{MobileViTV2 validation behaviour under the compute-matched training regime. Panel \textbf{(a)} shows validation loss as a function of optimizer step; panel \textbf{(b)} shows tumour-head validation accuracy over the same step axis. Solid curves are cubic-polynomial smoothed seed means, and shaded bands are estimated 95\% CI across ten seeds. The late-training bands are narrow because validation loss and tumour accuracy were highly stable across seeds. Unlike the CNN curves, these trajectories are not plotted against epochs because baseline and augmented MobileViTV2 runs were deliberately matched by update count. The curves stabilize rapidly and show only small final validation-accuracy separations, but the held-out paired tests show that those small differences were meaningful for the higher-capacity model. Because augmented batches contained fewer real images per update, comparable selected optimizer steps translated into substantially fewer real-data epochs before checkpoint selection (Table~\ref{tab:efficiency-stability}).}
  \label{fig:mobilevit-curves}
  \figcaprule
\end{figure}

\begin{figure*}[!tbp]
  \floatcaprule
  \centering
  \includegraphics[width=0.92\textwidth]{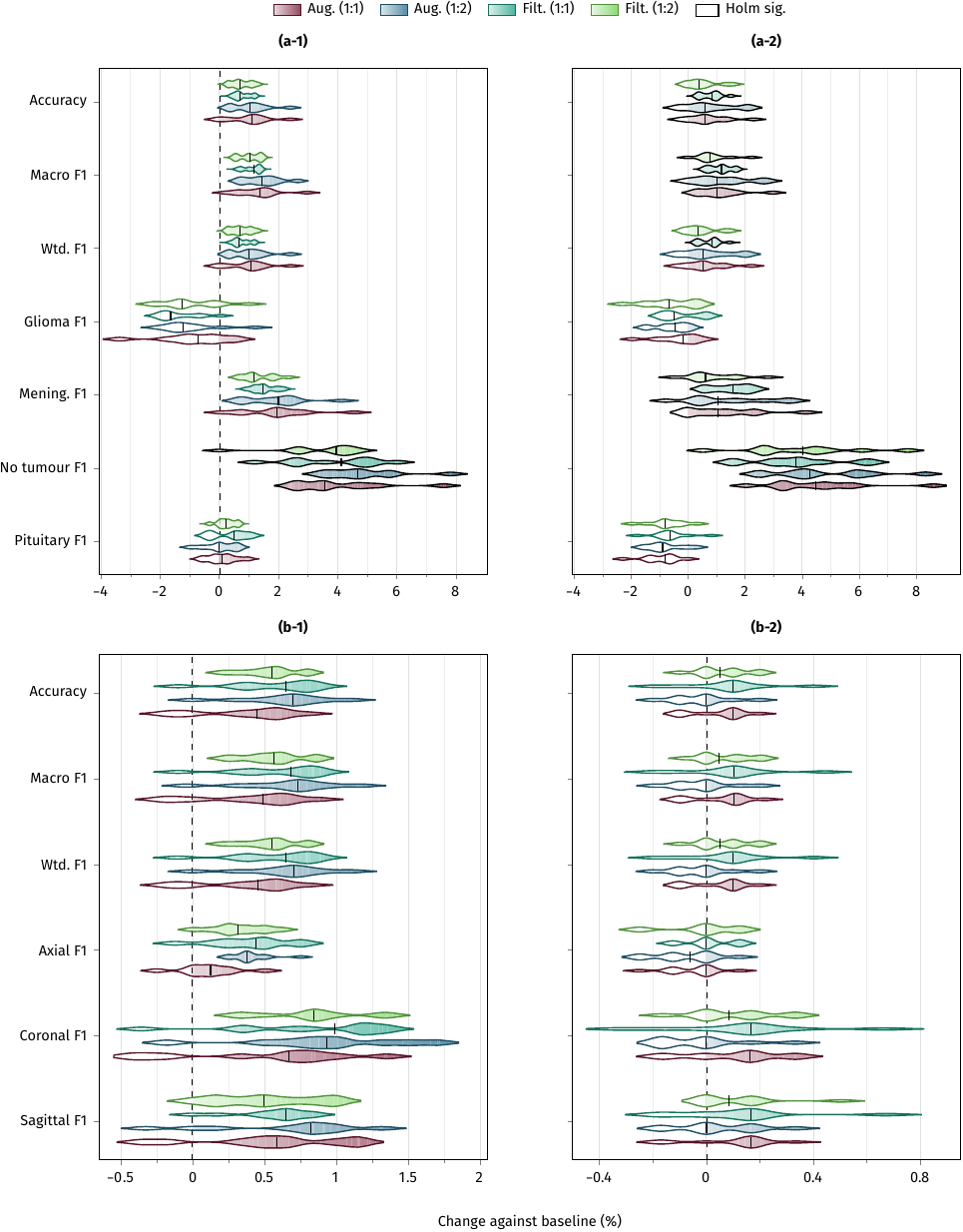}
  \caption{Paired downstream-performance changes for the compact CNN and MobileViTV2 relative to their real-only baselines. The four panels separate classifier and task: panel \textbf{(a-1)} shows CNN tumour metrics, panel \textbf{(a-2)} shows MobileViTV2 tumour metrics, panel \textbf{(b-1)} shows CNN anatomical-plane metrics, and panel \textbf{(b-2)} shows MobileViTV2 anatomical-plane metrics. Within each panel, rows are individual metrics and the horizontal axis is change against baseline in percent units; values to the right of the vertical zero line favour augmentation. Each horizontal violin corresponds to one augmentation condition. Negative portions are left unfilled, while positive portions use a zero-anchored colour gradient that starts very lightly at 0\% and intensifies as the positive change increases. Black-outlined violins mark comparisons that remained significant after Holm correction. The figure shows the architecture-dependent result in one view: CNN shifts are generally positive but statistically cautious, whereas MobileViTV2 shows the clearest corrected tumour-accuracy benefit, especially under filtered 1:1 augmentation.}
  \label{fig:cnn-mobilevit-deltas}
  \figcaprule
\end{figure*}

Taken together, these results suggest that synthetic data was most useful for the higher-capacity classifier when the synthetic distribution was not overrepresented. The filtered 1:1 condition struck the best balance---enough additional variation to help, but not enough synthetic volume to dominate the real data (Fig.~\ref{fig:mobilevit-curves}; Fig.~\ref{fig:cnn-mobilevit-deltas}). The baseline confusion matrices (Fig.~\ref{fig:mobilevit-confusion}) show MobileViTV2 matching the CNN on \textit{glioma}, \textit{no-tumour}, and \textit{pituitary} recall but with slightly lower \textit{meningioma} recall (83.9\% vs.\ 87.4\%); plane classification remained near-perfect.

Across all three classifier families, \textit{meningioma} was the least stable tumour class. The pattern is consistent with the feature-space view in Fig.~\ref{fig:umap-class}, where meningioma occupies a more diffuse region than the other tumour classes, and with the duplicate audit, in which removed overlap was disproportionately concentrated in meningioma axial images. A cautious reading is that meningioma presentation in this dataset is more heterogeneous and often boundary-adjacent, which makes it more sensitive to differences in skull stripping, crop placement, and feature extraction. That is precisely what makes the class informative: it is where synthetic augmentation is most likely to expose a distribution mismatch.

\begin{table*}[!t]
  \floatcaprule
\caption{Held-out tumour accuracy across classifier families and augmentation conditions. Each cell reports $\bar{x}$ \textpm{} $\sigma$ across ten seeds on the same 1,000-image held-out BRISC test set. Columns compare the real-only baseline against unfiltered synthetic augmentation at 1:1 and 1:2 ratios, plus feature-space-filtered augmentation at the same ratios. The final column gives the largest paired case-level tumour-accuracy delta against the real-only baseline in percent units, so it is not simply the difference between the displayed seed means. The column reports only the best tumour-accuracy change; harmful secondary effects are not summarized here, and the significant negative RF filtered 1:1 pituitary-F1 result is reported in Table~\ref{tab:primary-paired} and Fig.~\ref{fig:rf-deltas}.}
\label{tab:main-results}
\centering
\normalfont\plotfigurefont\scriptsize
\begin{tabularx}{\textwidth}{Yrrrrrr}
\toprule
Classifier & Real only & Synthetic 1:1 & Synthetic 1:2 & Filtered 1:1 & Filtered 1:2 & Best $\Delta$ (\%) \\
\midrule
RF on InceptionV3 pool3 features & \metriccell{80.69}{0.39} & \metriccell{80.30}{0.30} & \metriccell{80.32}{0.37} & \metriccell{79.59}{0.41} & \metriccell{80.29}{0.28} & -0.37 \\
Compact two-headed CNN & \metriccell{93.92}{0.45} & \metriccell{94.92}{0.66} & \metriccell{94.99}{0.60} & \metriccell{94.70}{0.31} & \metriccell{94.67}{0.47} & +0.72 \\
MobileViTV2 & \metriccell{94.18}{0.69} & \metriccell{94.91}{0.50} & \metriccell{94.96}{0.56} & \metriccell{94.98}{0.50} & \metriccell{94.70}{0.46} & +1.02 \\
\bottomrule
\end{tabularx}
\end{table*}

\begin{table*}[!b]
  \floatcaprule
\caption{Primary paired tumour-accuracy comparisons against the real-only baseline. Each row compares one classifier and one augmentation condition with that classifier's real-only baseline using paired case-level testing on the held-out BRISC test set. The change ($\Delta$) and confidence-interval columns report percent-unit deltas; positive values favour augmentation, negative values favour the real-only baseline. Holm correction is applied within each classifier's comparison family, defined as all four augmentation conditions crossed with the primary and secondary metrics (tumour accuracy, macro F1, weighted F1, per-class F1, and plane accuracy). Only the primary tumour-accuracy rows are shown; ``significant'' means significant after that family-wise correction.}
\label{tab:primary-paired}
\centering
\normalfont\plotfigurefont\scriptsize
\begin{tabularx}{\textwidth}{YYrrrY}
\toprule
Classifier & Condition & $\Delta$ (\%) & 95\% CI low & 95\% CI high & Holm result \\
\midrule
RF & Synthetic 1:1 & -0.39 & -1.18 & 0.36 & n.s. \\
RF & Synthetic 1:2 & -0.37 & -1.29 & 0.56 & n.s. \\
RF & Filtered 1:1 & -1.10 & -1.99 & -0.27 & n.s. (Holm) \\
RF & Filtered 1:2 & -0.40 & -1.38 & 0.57 & n.s. \\
CNN & Synthetic 1:1 & +0.57 & -0.11 & 1.25 & n.s. \\
CNN & Synthetic 1:2 & +0.72 & -0.02 & 1.49 & n.s. \\
CNN & Filtered 1:1 & +0.59 & -0.04 & 1.24 & n.s. \\
CNN & Filtered 1:2 & +0.63 & -0.04 & 1.33 & n.s. \\
MobileViTV2 & Synthetic 1:1 & +0.88 & 0.36 & 1.42 & significant \\
MobileViTV2 & Synthetic 1:2 & +0.96 & 0.38 & 1.58 & significant \\
MobileViTV2 & Filtered 1:1 & +1.02 & 0.54 & 1.54 & significant \\
MobileViTV2 & Filtered 1:2 & +0.77 & 0.27 & 1.32 & n.s. (Holm) \\
\bottomrule
\end{tabularx}
\end{table*}

\subsection{Training Efficiency and Seed Stability}
\label{sec:training-efficiency}

The held-out accuracy deltas alone understate one of the more practical effects of augmentation. In the compact CNN, every augmented condition reached its validation-loss-selected checkpoint substantially earlier than the real-only baseline (Table~\ref{tab:efficiency-stability}). The baseline selected epoch was 177.1~\textpm{}~24.5; the augmented checkpoints were selected at 99.1~\textpm{}~36.4 for unfiltered 1:1, 67.3~\textpm{}~21.4 for unfiltered 1:2, 102.0~\textpm{}~28.3 for filtered 1:1, and 63.3~\textpm{}~27.2 for filtered 1:2. The reductions span 42.4--64.3\% fewer epochs before the selected checkpoint, with exact paired sign-flip tests remaining below $p=0.006$ for all four comparisons. Even where the CNN accuracy gains were statistically cautious after Holm correction, augmentation consistently shifted the optimization trajectory: the model reached its best validation-loss state much sooner.

For the CNN, this is an epoch-level and real-pass statement, not a claim of lower wall-clock compute: each augmented epoch contains the real cohort plus additional synthetic images, so augmented epochs are heavier than real-only ones. The narrower but still practical point is that, under the same early-stopping rule, the CNN needed fewer repeated passes over the scarce real training cohort before reaching the validation-loss-selected checkpoint.

Likewise, MobileViTV2 showed the same pattern under a different measurement. Because MobileViTV2 runs were compute-matched by optimizer step, augmented conditions were not allowed to train for more updates than the baseline, and the selected optimizer step was consequently similar across conditions. The relevant gain was instead real-data exposure. The baseline selected checkpoint occurred after 53.2~\textpm{}~2.6 real-data epochs. In the 1:1 conditions it occurred after 26.6~\textpm{}~1.3 real-data epochs (a 50.0\% reduction); in the 1:2 conditions it occurred after 17.8~\textpm{}~1.1 real-data epochs (a 66.6\% reduction). The 1:2 selected checkpoints sat close to the fixed-step ceiling of about 18.7 real-data epochs, so the larger 1:2 percentage should be read partly as a consequence of the deliberate batch quota and step budget. Even so, the reduction is not merely cosmetic: the resulting held-out accuracies were maintained or improved, and MobileViTV2 seed-level tumour-accuracy standard deviation decreased in all four augmented conditions.

The seed-stability pattern was strongest for MobileViTV2 and more conditional for the other models. MobileViTV2 held-out tumour-accuracy standard deviation fell from 0.69\% at baseline to 0.50\%, 0.56\%, 0.50\%, and 0.46\% across the four augmented conditions, a relative reduction of 18.9--32.3\%. The CNN showed reduced dispersion only when filtering was applied at 1:1 (0.31\% vs.\ 0.45\% baseline); unfiltered CNN augmentation increased dispersion despite improving mean accuracy. The RF was mixed: three augmented conditions reduced seed-level standard deviation, while filtered 1:1 raised dispersion and lowered accuracy. The defensible conclusion is not that augmentation universally stabilizes every model, but that the higher-capacity MobileViTV2 benefited simultaneously in mean accuracy, seed stability, and real-data exposure efficiency.

\begin{table*}[!tbp]
  \floatcaprule
\caption{Training-efficiency and seed-stability summary for the two end-to-end classifiers. Accuracy is held-out tumour accuracy, reported as $\bar{x}$ \textpm{} $\sigma$ across ten seeds. For the CNN, selected-checkpoint effort is the epoch at which validation loss was minimized, measuring how many passes over the training schedule were needed before checkpoint selection. For MobileViTV2, selected-checkpoint effort is expressed as real-data epochs seen by the selected optimizer step, since MobileViTV2 runs were compute-matched by update count and augmented batches contained fewer real images. \textbf{Sign conventions differ between the two delta/effort columns:} for Seed-$\sigma$ $\Delta$, negative values indicate reduced seed-to-seed dispersion; for Reduction, positive values indicate fewer selected-checkpoint or real-data epochs than baseline. The sign-flip $p_{\mathrm{sf}}$ is an exact paired seed-level test of whether selected-checkpoint effort was consistently lower than baseline; with ten paired seeds, values reported as 0.002 are the rounded minimum attainable two-sided value.}
\label{tab:efficiency-stability}
\centering
\normalfont\plotfigurefont\scriptsize
\begin{tabularx}{\textwidth}{YYrrrrr}
\toprule
Model & Condition & Accuracy (\%) & Seed-$\sigma$ $\Delta$ (\%) & Selected checkpoint effort & Reduction (\%) & $p_{\mathrm{sf}}$ \\
\midrule
CNN & Baseline & \metriccell{93.92}{0.45} & -- & \metriccell{177.1}{24.5} epochs & -- & -- \\
CNN & Synthetic 1:1 & \metriccell{94.92}{0.66} & +44.5 & \metriccell{99.1}{36.4} epochs & 44.0 & 0.0059 \\
CNN & Synthetic 1:2 & \metriccell{94.99}{0.60} & +32.3 & \metriccell{67.3}{21.4} epochs & 62.0 & 0.0020 \\
CNN & Filtered 1:1 & \metriccell{94.70}{0.31} & -31.1 & \metriccell{102.0}{28.3} epochs & 42.4 & 0.0020 \\
CNN & Filtered 1:2 & \metriccell{94.67}{0.47} & +2.8 & \metriccell{63.3}{27.2} epochs & 64.3 & 0.0020 \\
\midrule
MobileViTV2 & Baseline & \metriccell{94.18}{0.69} & -- & \metriccell{53.2}{2.6} real epochs & -- & -- \\
MobileViTV2 & Synthetic 1:1 & \metriccell{94.91}{0.50} & -26.8 & \metriccell{26.6}{1.3} real epochs & 50.0 & 0.0020 \\
MobileViTV2 & Synthetic 1:2 & \metriccell{94.96}{0.56} & -18.9 & \metriccell{17.8}{1.1} real epochs & 66.6 & 0.0020 \\
MobileViTV2 & Filtered 1:1 & \metriccell{94.98}{0.50} & -27.7 & \metriccell{26.6}{1.3} real epochs & 50.0 & 0.0020 \\
MobileViTV2 & Filtered 1:2 & \metriccell{94.70}{0.46} & -32.3 & \metriccell{17.8}{1.1} real epochs & 66.6 & 0.0020 \\
\bottomrule
\end{tabularx}
\end{table*}

\subsection{Effect of Ratio and Filtering}

Notably, the 1:2 ratio was not uniformly better than 1:1. In the random forest, both unfiltered ratios underperformed baseline and filtering did not rescue the model. In the CNN, 1:2 produced the highest mean accuracy, but without corrected significance for the primary endpoint. In MobileViTV2, unfiltered 1:2 was significant, but filtered 1:1 achieved the strongest corrected primary result and the highest mean accuracy. More synthetic data, then, is not automatically better.

Filtering had similarly mixed effects. It was harmful in the RF filtered 1:1 condition, did not clearly improve the CNN primary endpoint, and was most useful for MobileViTV2 at 1:1. The filter is thus architecture- and ratio-dependent: it likely removed some harmful outliers, but it may also have removed variation that fixed-feature classifiers such as the RF could have used.

\section{Discussion}

\subsection{Interpretation of the Main Findings}
\label{sec:discussion-main}

The headline result is not that StyleGAN2-ADA augmentation always improves BRISC classification. It does not. The random forest did not benefit, and some filtered RF outcomes worsened; the CNN showed consistent but statistically cautious gains; MobileViTV2 provided the strongest evidence that synthetic augmentation can help in this constrained setting, especially with filtered 1:1 augmentation.

Nonetheless, augmentation can be valuable when final accuracy gains are modest. Across all augmented CNN conditions, the validation-loss-selected checkpoint occurred 42--64\% earlier than in the real-only baseline. In MobileViTV2, compute-matched augmented runs reached their selected checkpoints after 50--67\% fewer real-data epochs. In constrained medical settings, where real examples are the scarce resource, synthetic data can reduce how many repeated passes over the real cohort are needed to reach a useful model state, even when the final held-out accuracy improvement is small.

The pattern rules out the simple version of the synthetic-data argument: adding GAN images is not equivalent to adding real data. The value of synthetic augmentation depends on generator coverage, the augmentation ratio, the filtering rule, the downstream architecture, and how easily the classifier can ignore or exploit artifacts. The architecture-dependent response is consistent with a bias--variance reading. The RF operates on frozen features and so cannot reduce bias from additional training examples that carry domain-specific information absent from the pretrained feature space. The end-to-end models, by contrast, can adjust their internal representations to extract signal from synthetic variation, but they also risk increased variance when synthetic images introduce distributional noise. Direct CNN-versus-MobileViTV2 comparisons should be read with the batch protocol in mind: the CNN used a shuffled appended manifest, whereas MobileViTV2 enforced per-step real/synthetic quotas to keep compute matched. The filtered 1:1 MobileViTV2 condition may sit at a favourable operating point---one where synthetic diversity reduces bias (by expanding the effective training distribution) without substantially increasing variance, since filtering removes the most atypical samples and the 1:1 ratio prevents synthetic data from dominating the loss landscape.

\subsection{Why Synthetic Augmentation May Help}

In principle, synthetic images can improve generalization when they add plausible variation near the real distribution. In a class-plane partition with only a few hundred real images, even imperfect synthetic examples may dampen overfitting to repeated framing, contrast, or tumour-appearance patterns. The faster CNN checkpoint selection hints at a related optimization mechanism: synthetic examples may make the supervised loss landscape less dependent on the idiosyncratic ordering of real samples, letting the model reach a stable validation state earlier. MobileViTV2 may have benefited because its pretrained hybrid representation, compute-matched training, and Mixup regularization jointly equipped it to absorb useful synthetic variation without overfitting to source-specific artifacts. Mixup, by interpolating training pairs and their labels, produces convex combinations that smooth decision boundaries and reduce the classifier's sensitivity to individual training examples. Where synthetic images carry subtle distributional differences from real data, that smoothing may prevent the model from memorizing source-identifying shortcuts that a non-regularized classifier would happily exploit.

In particular, the strongest evidence for this mechanism is the filtered 1:1 MobileViTV2 result. The filter likely discarded the most atypical synthetic images while preserving enough diversity to expand the training distribution. The gain is modest in absolute terms, but it is meaningful because it was measured on held-out real test data and remained significant after family-wise correction. Just as important, the same condition reduced seed-to-seed tumour-accuracy standard deviation by 27.7\% and selected its checkpoint after half as many real-data epochs as the baseline.

\subsection{Why Synthetic Augmentation May Hurt}

Conversely, synthetic augmentation can hurt when generated images carry artifacts, narrow coverage, unrealistic anatomical boundaries, or feature distributions that diverge from the real test set. A classifier can then learn source-specific signals instead of tumour-relevant features, and the risk grows as the synthetic ratio rises high enough to shift the training distribution away from the real one.

The random forest results illustrate that risk most clearly. The RF operates on fixed InceptionV3 features and cannot learn a medical-domain representation from the raw images. When synthetic images occupy slightly different feature regions, adding them blurs decision boundaries instead of sharpening them. The filtered 1:1 RF condition shows that feature-space typicality is not automatically aligned with classifier utility. The practical implication is that practitioners who select synthetic training data by feature-space quality should verify that the criterion matches the downstream classifier's decision surface, not merely the generator's output distribution.

\subsection{Importance of Class-Plane Generators}

Training twelve separate generators was central to the experiment. After all, brain MRI is not a single homogeneous image distribution: tumour class and anatomical plane jointly affect anatomy, texture, contrast, lesion presentation, and spatial structure. A class-plane generator has a narrower task and can thus allocate its capacity more efficiently than a single generator trained across all classes and planes.

Even so, the cost is reduced sample size. Some generators were trained on only a few hundred images, which limits distributional coverage and likely contributes to the low recall values reported above. The contribution here is to test that trade-off honestly rather than pretend it does not exist.

\subsection{Limitations}

Several limitations bound the scope of these conclusions. The study uses a single public dataset, and any dataset-specific bias that survives the BRISC curation process \cite{fateh2026brisc} can affect both baseline and augmented results. The overlap audit removed exact image duplication and documented the removed class-plane distribution, but no subject identifiers were available, so the final split is image-level non-overlapping rather than patient-level independent. Residual related-scan structure, repeated acquisitions, or slices from the same subject could remain across the public split and inflate every classifier, the real-only baselines included. This matters because BRISC is assembled from multiple public sources; the reported accuracies should be read as held-out image-level performance on the BRISC split rather than external patient-level validation. The 356 close perceptual-hash neighbours and 5 close Inception-feature neighbours were flagged but not removed, and no sensitivity analysis dropping those near-neighbour test cases was performed.

The images are two-dimensional preprocessed slices at 128~$\times$~128 pixels. This resolution made twelve generator runs and multi-seed downstream testing feasible on a single workstation, but it is not a substitute for full-resolution or volumetric analysis. It also constrains the VLM blind test: GPT-5.5 saw the same preprocessed 128~$\times$~128 representations used by the classifiers, not the original clinical-resolution MRIs. The VLM result should be read as evidence of surface-level realism at the working resolution. It does not establish diagnostic plausibility, radiologist-level indistinguishability, or the absence of subtle frequency artifacts that may surface at higher resolution. The GPT-5.5 audit is also a computational surrogate for human review; no radiologist or clinical-expert evaluation was performed.

On the generative side, synthetic images were evaluated through generic InceptionV3 feature spaces rather than domain-specific encoders or radiologist assessment; FID, KID, precision, and recall should not, for that reason, be taken as measures of diagnostic plausibility. The same feature dependency affects the RF, which is best understood as a fixed-representation baseline rather than a competitive medical classifier. The study also did not run a full-ADA-versus-restricted-ADA ablation, did not train downstream classifiers under alternative Mahalanobis cutoffs, and does not include a figure of rejected synthetic failure modes. Those omissions matter because the restricted ADA subset, the $\rho_{\mathrm{pool}}=2.5$ pool multiplier, and the $q_{0.975}$ filtering threshold are plausible design choices, not optimized hyperparameters.

The training-efficiency analysis is secondary and internal to the training protocol. Earlier selected CNN epochs and fewer MobileViTV2 real-data epochs are useful operational signals, especially when real images are scarce, but they do not substitute for external validation or a complete compute-cost analysis. The final learned multitask log-variance values $s_t$ and $s_p$ were used during optimization but were not serialized in the saved seed outputs, so this manuscript cannot compare how augmentation changed the learned tumour-versus-plane weighting. In addition, the local PyTorch build was a pre-release nightly; although all scripts and result artifacts record the software stack, exact bitwise replication may depend on access to the same dated wheel and CUDA environment. Finally, the strongest accuracy gains were small in absolute terms: the MobileViTV2 filtered 1:1 improvement of roughly 1\% absolute warrants replication on additional cohorts before informing deployment decisions.

\subsection{Future Work}

The most pressing extension is replication on additional MRI cohorts with external held-out sets and, where metadata allows, verified patient-level independence. Radiologist assessment of synthetic images would anchor visual plausibility to diagnostic rather than statistical criteria. A direct comparison with diffusion models under identical class-plane constraints would clarify whether the moderate recall coverage seen here is a GAN-specific limitation or simply a general consequence of training on a few hundred images. On the filtering side, domain-specific encoders, uncertainty-aware candidate scoring, and classifier-disagreement filtering are plausible alternatives to the InceptionV3 Mahalanobis approach used here. In any follow-on study, generator-quality metrics should remain diagnostic aids rather than primary selection criteria; the downstream held-out result is the only measure that directly answers the utility question.

\section{Conclusion}

Twelve class-plane-specific StyleGAN2-ADA generators were trained on constrained BRISC partitions, synthetic images were generated, optional feature-space filtering was applied, and augmentation utility was evaluated across a random forest, a convolutional network, and MobileViTV2.

The answer to the utility question is not a simple yes or no. Synthetic augmentation did not improve the random forest, produced consistent CNN mean-level shifts that did not survive Holm correction, and gave the clearest benefit to MobileViTV2 under filtered 1:1 augmentation: +1.02\% absolute tumour accuracy (95\% CI: 0.54--1.54\%; Holm-corrected $p=0.0104$). The training-history analysis adds a second practical conclusion: synthetic data can improve training efficiency even when final accuracy gains are modest. In the CNN, every augmented condition reached its selected checkpoint 42--64\% earlier than baseline; in MobileViTV2, augmented conditions reached their selected checkpoints after 50--67\% fewer real-data epochs while preserving or improving held-out accuracy. GAN-based augmentation in medical imaging is architecture- and ratio-sensitive, and its value materialises only when generator coverage, filtering strategy, synthetic ratio, and downstream model capacity align. Volume of synthetic data is not a substitute for that alignment; well-aligned synthetic data, however, can ease the pressure on scarce real examples.

The negative and mixed outcomes are part of the result, not a footnote to it. They show why synthetic medical-image augmentation must be judged by held-out task performance, not by visual realism alone. StyleGAN2-ADA can help under selected conditions, but it is not a substitute for real annotated MRI or clinical validation. Future studies should accordingly report both negative and positive outcomes, since the boundary between useful augmentation and harmful distribution shift is itself the object of interest.

\section*{Data and Code Availability}

The BRISC 2025 dataset is publicly available through the official Kaggle release cited in this manuscript. The public project repository for this study, at \monourl{https://github.com/joriegac/brisc2025-stylegan2-augmentation}, is the release point for the manuscript source and the preprocessing, \texttt{StyleGAN2-ADA} launch, synthetic-generation, filtering, classifier-training, statistical-analysis, and plotting scripts, so that the numerical results can be audited from the same intermediate artifacts. Figures and tables were generated from \texttt{JSONL}, \texttt{CSV}, and \texttt{manifest} artifacts stored with the project: \texttt{plot-registry} \texttt{JSONL} files for the \texttt{PGFPlots} figures, classifier seed outputs, training-history \texttt{JSONL} files, filtering audits, and \texttt{VLM} batch outputs. The \texttt{VLM} audit artifacts retain \texttt{model\_string}, \texttt{request\_grouping}, \texttt{parsed\_outputs}, \texttt{blinded\_image\_index}, and \texttt{private\_answer\_key}; no immutable API-side \texttt{model\_snapshot\_id} was available. The synthetic image pools are large; they can be regenerated from the released scripts, selected checkpoints, seeds, and manifests rather than distributed as mandatory figure assets.

\end{document}